\documentclass[useAMS,usenatbib]{mn2e}
\usepackage[T1]{fontenc}
\usepackage{aecompl}

\usepackage{amsmath,enumitem,booktabs,fixltx2e,mdframed}
\usepackage{psfig,graphicx,threeparttable}
\newcommand{\tsu}[1]{\textsuperscript{#1}}
\newcommand{\tsb}[1]{\textsubscript{#1}}
\graphicspath{{graphics/}}
\usepackage{xfrac}
\usepackage{float}

\newcount\longrefs
\def\aap{\ifnum\longrefs=1 {Astron.\ Astrophys.}\else 
                           {A\hbox{\rm \&}A}\fi}
\def\aapr{\ifnum\longrefs=1 {Astron.\ Astrophys.\ Rev.}\else 
                            {A\hbox{\rm \&}AR}\fi}
\def\aaps{\ifnum\longrefs=1 {Astron.\ Astrophys.\ Suppl.}\else 
                            {A\hbox{\rm \&}A Suppl.}\fi}
\def\aipcs{\ifnum\longrefs=1 {Am.\ Inst.\ Phys.\ Conf.\ Series}\else
                             {AIP Conf.\ Ser.}\fi}
\def\aj{\ifnum\longrefs=1 {Astron.\ J.}\else 
                          {AJ}\fi} 
\def\ao{\ifnum\longrefs=1 {Applied Optics}\else 
                           {Appl.\ Opt.}\fi} 
\def\aspcs{\ifnum\longrefs=1 {Astron.\ Soc.\ Pacific Conf.\ Series}\else 
                           {ASP Conf.\ Ser.}\fi} 
\def\apj{\ifnum\longrefs=1 {Astrophys.\ J.}\else 
                           {ApJ}\fi} 
\def\apjl{\ifnum\longrefs=1 {Astrophys.\ J. Lett.}\else 
                            {ApJ}\fi} 
\def\aplett{\ifnum\longrefs=1 {Astrophys.\ J. Lett.}\else 
                            {ApJ}\fi} 
\def\apjs{\ifnum\longrefs=1 {Astrophys.\ J. Suppl.}\else 
                            {ApJS}\fi}
\def\apss{\ifnum\longrefs=1 {Astrophys.\ and Space Science}\else 
                            {Astrophys.\ Space Sci.}\fi}
\def\araa{\ifnum\longrefs=1 {Ann.\ Rev.\ Astron.\ Astrophys.}\else 
                            {ARA\hbox{\rm \&}A}\fi}
\def\azh{\ifnum\longrefs=1 {Astronomicheskii Zhurnal}\else 
                            {Astron.\ Zhur.}\fi}
\def\baas{\ifnum\longrefs=1 {Bull.\ Am.\ Astron.\ Soc.}\else 
                            {BAAS}\fi}
\def\bain{\ifnum\longrefs=1 {Bull.\ Astronom.\ Institutes Netherlands}\else
                            {Bull.\ Astr.\ Inst.\ Neth.}\fi}
\def\gca{\ifnum\longrefs=1 {Geochim.\ Cosmochim.\ Acta}\else 
                           {Geochim.\ Cosmochim.\ Acta}\fi}
\def\grl{\ifnum\longrefs=1 {Geophys.\ Res.\ Lett.}\else 
                           {Geoph.\ Res.\ Lett.}\fi}
\def\iaucirc{\ifnum\longrefs=1 {IAU Circulars}\else 
                          {IAU Circ.}\fi}
\def\ip{\ifnum\longrefs=1 {in press}\else 
                          {in press}\fi}
\def\jgr{\ifnum\longrefs=1 {J.\ Geophys.\ Res.}\else 
                           {J.\ Geophys.\ Res.}\fi}  
\def\jrasc{\ifnum\longrefs=1 {J.\ Royal Astron.\ Soc.\ Canada}\else 
                           {JRAS Can.}\fi}  
\def\memsai{\ifnum\longrefs=1 {Mem.~Soc.~Astron.~Italiana}\else
                              {MemSAI}\fi}
\def\mnras{\ifnum\longrefs=1 {Mon.\ Not.\ Roy.\ Astron.\ Soc.}\else 
                             {MNRAS}\fi} 
\def\nat{\ifnum\longrefs=1 {Nature}\else 
                           {Nat}\fi}
\def\pasj{\ifnum\longrefs=1 {Pub.\ Astron.\ Soc.\ Japan}\else 
                            {PASJ}\fi} 
\def\pasp{\ifnum\longrefs=1 {Pub.\ Astron.\ Soc.\ Pacific}\else 
                            {PASP}\fi} 
\def\physscr{\ifnum\longrefs=1 {Physica Scripta}\else 
                            {Phys.\ Scrip.}\fi} 
\def\planss{\ifnum\longrefs=1 {Planetary \& Space Science}\else 
                            {Plan. \& Space Sci.}\fi} 
\def\procspie{\ifnum\longrefs=1 {Proc.\ SPIE}\else 
                            {Proc.\ SPIE}\fi} 
\def\qjras{\ifnum\longrefs=1 {Quarterly J.\ Royal Astron.\ Soc.}\else 
                            {QJRAS}\fi} 
\def\sa{\ifnum\longrefs=1 {Soviet Astron..}\else 
                               {Sov.\ Astron.}\fi}
\def\skytel{\ifnum\longrefs=1 {Sky \& Telescope}\else 
                            {Sky \& Tel.}\fi} 
\def\solphys{\ifnum\longrefs=1 {Solar Phys.}\else 
                               {Sol.\ Phys.}\fi}
\def\ssr{\ifnum\longrefs=1 {Space Science Rev.}\else 
                               {Space\ Sci.\ Rev.}\fi}
\def\zap{\ifnum\longrefs=1 {Zeitschr.\ f.\ Astrophysik}\else
                               {Z.\ Astrophys.}\fi}

\title[Roche tomography of CVs - VI. Differential rotation of AE Aqr]{Roche tomography of cataclysmic variables - VI. \\Differential rotation of AE Aqr - Not tidally locked!}
\author[C.A. Hill et  al.]{C.A. Hill$^{1}$\thanks{E-mail:
chill17@qub.ac.uk}, C.A. Watson$^{1}$, T. Shahbaz$^{2,3}$, D. Steeghs$^4$, and V.S. Dhillon$^5$\\
$^1$Astrophysics Research Centre, Queen's University Belfast, Belfast, BT7 1NN, Northern Ireland, UK\\
$^2$Instituto de Astrof\'{i}sica de Canarias (IAC), E-38200 La Laguna, Tenerife, Spain \\
$^3$Departamento de Astrof\'{i}sica, Universidad de La Laguna (ULL), E-38206 La Laguna, Tenerife, Spain\\
$^4$Department of Physics, University of Warwick, Coventry, CV4 7AL, UK\\
$^5$Department of Physics \& Astronomy, University of Sheffield, Sheffield, S3 7RH, UK}
\begin{document}

\date{}

\pagerange{\pageref{firstpage}--\pageref{lastpage}} \pubyear{2013}

\maketitle

\label{firstpage}

\begin{abstract}
We present Roche tomograms of the K4V secondary star in the cataclysmic variable AE Aqr, reconstructed from two datasets taken 9 days apart, and measure the differential rotation of the stellar surface. The tomograms show many large, cool starspots, including a large high-latitude spot and a prominent appendage down the trailing hemisphere. We find two distinct bands of spots around $22^{\circ}$ and $43^{\circ}$ latitude, and estimate a spot coverage of 15.4-17 per cent on the northern hemisphere. 

Assuming a solar-like differential rotation law, the differential rotation of AE Aqr was measured using two different techniques. The first method yields an equator-pole lap time of 269 d and the second yields a lap time of 262 d. This shows the star is not fully tidally locked, as was previously assumed for CVs, but has a co-rotation latitude of $\sim40^{\circ}$. We discuss the implications that these observations have on stellar dynamo theory, as well as the impact that spot traversal across the L\tsb{1} point may have on accretion rates in CVs as well as some of their other observed properties.

The entropy landscape technique was applied to determine the system parameters of AE Aqr. For the two independent datasets we find $M_{1} = 1.20$ and $1.17~\mathrm{M}_{\odot}$, $M_2 = 0.81$ and $0.78~\mathrm{M}_{\odot}$, and orbital inclinations of $50^{\circ}$ to $51^{\circ}$ at optimal systemic velocities of $\gamma = -64.7$  and $-62.9$~kms\tsu{-1}.
\end{abstract}


\section{Introduction}
\label{sec:intro}
Differential rotation of a solar-like star is thought to be a key component in stellar dynamo theory, converting and amplifying poloidal fields into toroidal fields. The magnetic activity observed on these stars is assumed to be based on a dynamo mechanism operating in the outer convection zone. The details of the generation and amplification mechanisms are currently not well understood due to the complex interactions involved and the lack of observational constraints. To test and better constrain stellar dynamo theory, we can compare models with observations of differential rotation (DR) on stars of different masses and rotation rates.

Previous measurements of DR on single stars include Donati and Cameron~\citeyearpar{donati1997abdor} in their study of the rapidly rotating single K-type star AB Dor ($\mathrm{P_{rot}} = 12.36$ h). They found an equator-pole lap time of 110 d, close to solar value of 120 d. However, this displayed significant variability over several years, falling from 140 d to 70 d in the period 1988--1992, when the DR rate doubled. The Lupus post T Tauri star RX J1508.6±4423 ($\mathrm{P_{rot}} = 7.4$ h) was found by \cite{donati2000} to have a lap time of $50\pm10$ d, and the late-type single star PZ Tel ($\mathrm{P_{rot}} = 22.7$ h) was found by \cite{barnes2000} to have a lap time of 72--100 d. However, in the less rapidly rotating K2 dwarf star LQ Hya ($\mathrm{P_{orb}} = 38.4$ h), \cite{kovari2005} found it to be much more rigidly rotating, with a lap time of $\sim280$ d.

Work by Scharlemann~\citeyearpar{scharlemann1982} suggested DR should be fairly suppressed in tidally locked systems such as CVs, and this was found in the tidally-locked pre-CV V471 Tau ($\mathrm{P_{rot}} = 12.5$ h) by Hussain et al.~\citeyearpar{hussain2006}. They found the surface shear consistent with solid body rotation, concluding that tidal locking may inhibit DR, but that this reduced shear does not affect the overall magnetic activity levels in this active K dwarf. Similarly, \cite{petit2004} found weak DR on the RS CVn system HR 1099 ($\mathrm{P_{rot}}\sim\mathrm{P_{orb}} = 68.1$ h) with a lap time of $\sim480$ d. 

It is clear that different systems display a variety of DR rates, and so the study and comparison of stars of varying fundamental parameters is crucial for our understanding of the underlying dynamo mechanism. To this end, a measure of DR on cataclysmic variables would allow a test of dynamo theory in a parameter space with both rapid rotation and tidal distortion. 

In addition, magnetic activity plays a critical role in the evolution of CVs, driving them to shorter orbital periods via magnetic braking (e.g. \citealt{kraft1967,mestel1968,spruit1983,rappaport1983}), with the transition of secondary to a fully convective state --- supposedly shutting down magnetic activity --- invoked to explain the period gap and dearth of CVs with 2-3 h periods. On shorter timescales, magnetic activity has been invoked to explain variations in CV orbital periods, mean brightnesses, mean outburst durations and outburst shapes (e.g. \citealt{bianchini1990,richman1994,ak2001}). In a study of the polar-type CV AM Her, Livio and Pringle~\citeyearpar{livio1994} suggested that mass-transfer variations were due to starspots traversing the L\tsb{1} point. This suggestion is supported by the more recent work of Hessman et al.~\citeyearpar{hessman2000} in their derivation of the mass transfer history of AM Her. They found that spot traversal across the L\tsb{1} point is a valid mechanism for mass-transfer variation, if magnetic flux is preferentially produced around the L\tsb{1} point or if there is a mechanism which forces spot groups appearing at higher latitudes to wander down towards the L\tsb{1} point. Indeed, \cite{holzwarth2003} suggest tidal distortions may force starspots to form at preferred longitudes.

Clearly magnetic activity plays an important role in understanding the evolution and behaviour of CVs, and since differential rotation is a key component of such activity, a study of the secondary star in CVs allows a test of stellar dynamo theory and further insight into this class of system. In this paper we present intensity maps of the stellar surface on the secondary star in the CV AE Aqr using Roche tomography (see \cite{watson2001}; \cite{dhillon2001}; \cite{watson2003} for details), and measure the stellar surface differential rotation rate.

\section{Observations and reduction}
Spectroscopic observations were carried out over 4 half-nights on 2009 August 27--28 and September 5--6. From here on we refer to nights 1 and 2 as block 1, and nights 3 and 4 as block 2. The data were acquired using the 8 m VLT, situated at Cerro Paranal in Chile. The spectroscopic observations of AE Aqr were carried out using the Ultraviolet and Visual Echelle Spectrograph (UVES). A thinned EEV CCD-44 chip with 2K x 4K pixels was used in the blue channel, and a mosaic of an EEV CCD-44 and a MIT/LL CCID-20 chip (both 2K x 4K) were used in the red channel. UVES was used in the Dichroic-1/Standard setting (390+580 nm) mode, allowing a wavelength coverage of 3259 \r{A}--4563 \r{A} in the blue arm and 4726 \r{A}--6835 \r{A} in the red arm. With a slit width of 0.9 arcsec, a spectral resolution of around 46,000 ($\sim6.52$ kms$^{-1}$) was obtained in the blue, and a spectral resolution of around 43,000 ($\sim6.98$ kms$^{-1}$) was obtained in the red, with the chip binning left as 1x1. 

The blue and red spectra were taken quasi-simultaneously. During the first observing block, the blue and red spectra were taken using 228 s and 230 s exposures respectively, in order to minimise velocity smearing of the data due to the orbital motion of the secondary star. Each exposure corresponds to 0.65\% of the orbit.
%
%
%
Different exposure times were used in an attempt to compensate for the longer readout time of the blue CCD. The exposures were taken in blocks of 10, allowing red and blue data to remain quasi-simultaneous. For the second observing block, red and blue spectra were taken using 230 s exposures. For both observing blocks comparison ThAr lamp exposures were taken at the start and end of the night for the purpose of wavelength calibration. In the first block with this setup we obtained 118 useable spectra. 12 red spectra were lost due to a temporary shutter malfunction. In addition, spectral type templates were obtained for the purpose of determining the orbital phasing of AE Aqr via cross-correlation of absorption lines.

The seeing was around 2 arcsec at the start of night one, improving to 1.2 arcsec by mid observation, with seeing on the second night starting out around 4 arcsec, greatly improving to 1--0.9 arcsec by the middle of the night before worsening towards the end. In the second block (9 days later) we obtained 121 useable spectra in both the blue and red, giving a 100 per cent orbit coverage of AE Aqr. The seeing was 1.2--0.8 arcsec on the first night and 0.4--0.8 arcsec the second night. Table~\ref{tab:obs_data} gives a journal of the observations.

\begin{table*}
\caption{Log of the VLT spectroscopic observations of AE Aqr, the spectral-type template stars and flux standards. The first column gives the object name, columns 2--4 list the UT Date and the exposure start and end times, respectively. Columns 5--6 list the exposure times and number of spectra taken for each object and the final column indicates the type of science frame taken.} 
\label{tab:obs_data}
\begin{tabular}{lllllll}
\toprule
Object & UT Date & UT Start & UT End & T\tsb{exp}(s) & No. spectra & Comments \\
\midrule
HD 187760	&	2009 Aug 27	&	23:15	&	23:19	&	240	&	1	&	K4V template\\
HD 187760	&				&	23:22	&	23:30	&	480	&	1	&	\\
GJ 1247		&				&	23:57	&	00:03	&	360	&	1	&	K3V template\\
GL 653		&				&	23:40	&	23:44	&	240	&	1	&	K5V template\\
AE Aqr		&				&	23:49	&	05:11	&	230	&	57	&	Target spectra\\
GD50		&	2009 Aug 28	&	10:10	&	10:25	&	900	&	1	&	Flux standard\\
GJ 1192		&				&	23:15	&	23:22	&	360	&	1	&	K3V template\\
GL 653		&				&	23:42	&	23:46	&	240	&	1	&	K5V template\\
GJ 1247		&				&	23:57	&	00:03	&	360	&	1	&	K3V template\\
AE Aqr		&	2009 Aug 29	&	00:08	&	04:55	&	230	&	61	&	Target spectra\\
HR 7596		&	2009 Sept 05	&	23:31	&	23:31	&	4	&	1	&	Flux standard\\
HR 7596		&				&	23:34	&	23:34	&	15	&	1	&	\\
HR 7596		&				&	23:36	&	23:37	&	30	&	1	&	\\
AE Aqr		&			 	&	23:43	&	04:38	&	230	&	61	&	Target spectra\\
AE Aqr		&	2009 Sept 07	&	00:34	&	05:20	&	230	&	60	&	Target spectra\\
\bottomrule
\end{tabular}
\end{table*}


\subsection{Data reduction}
\label{sec:datareduction}
The raw data were reduced using the ESO UVES pipeline. This automatically processes all calibration frames and then conducts bias subtraction and flat fielding, wavelength calibration, sky subtraction, flux calibration and optimal extraction of the target frames. The bias frames used were those closest in time to the science data. The final output consists of 1-d spectra for the blue and red arms.

It should be noted that Roche tomography cannot be performed on data which has not been corrected for slit losses. This is because the variable contribution of the secondary star to the total light of a CV forces one to use relative line fluxes during the mapping process and prohibits the usual method of normalising the spectra. We discuss our corrections for slit losses in section~\ref{sec:scaling}.

Over both observing blocks the peak signal-to-noise of the blue spectra ranged from 5--31 (typically $\sim17$), and from 17--64 (typically $\sim46$) in the red.

\section{Ephemeris, Radial velocity curves and Continuum fitting}

\subsection{Ephemeris and Radial velocity curve}
\label{sec:ephemeris}
In order to construct an artefact-free Roche tomogram the orbital ephemeris is required. A new ephemeris for AE Aqr was determined by cross-correlation with a template star of spectral type K4V. We only considered the spectral region 6000--6500 \r{A} for simplicity, as it contains strong absorption lines from the secondary star and reduces the probability of introducing a sloping continuum contribution from the blue accretion regions. The AE Aqr and K4V template spectra were first normalised by dividing by a constant, and the continuum was then subtracted off using a third order polynomial fit, thus preserving line strength in the spectral region of interest. 

In order to perform the cross-correlation, the template spectrum was artificially broadened by an arbitrary amount to account for the rotational velocity ($v\sin{i}$) of the secondary star. The AE Aqr spectra were then cross-correlated with the artificially broadened template giving an initial estimate of the radial velocity of the secondary star at that orbital phase. In order to derive an improved estimate of the rotational broadening of the secondary star, the orbital motion was subtracted. These spectra were then averaged to provide one high signal-to-noise orbitally-corrected spectrum. A new value for the rotational broadening of AE Aqr was then obtained by artificially broadening the template spectrum in 0.1 kms\tsu{-1} steps and optimally subtracting the broadened template from the orbitally corrected AE Aqr spectrum. A new $v\sin{i}$ value was then obtained by adopting the broadening that needed to be applied to the template spectrum in order to minimise the residuals in the optimal subtraction.  This new broadening value was then applied to the template and the whole process repeated until the rotational broadening value no longer changed. This typically took 3 iterations. The whole procedure was carried out separately for block 1 and block 2 data.

\begin{figure}
\centering
\includegraphics[width=0.5\textwidth]{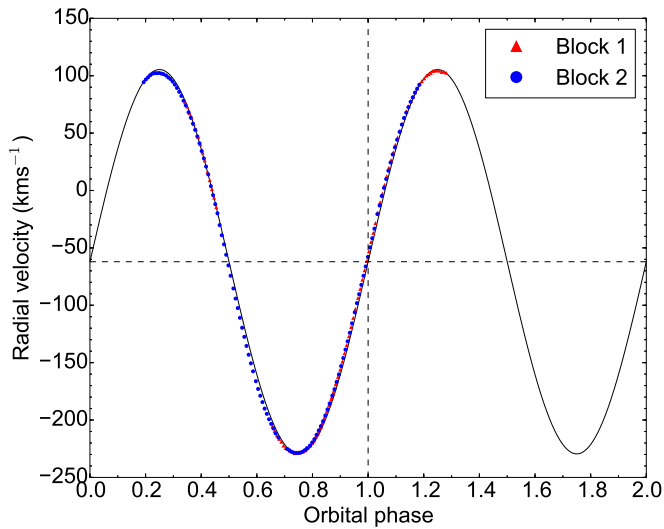}
\caption{Radial velocity curve of AE Aqr, which has been phase-folded for clarity. The solid line is a least-squares sinusoid fit to the RV points, assuming a circular orbit. Triangles are block 1 data, circles are block 2 data. The error bars are smaller than the points in this figure.}
\label{fig:rvcurve}
\end{figure}

Through the above process a cross-correlation function (CCF) was calculated for each AE Aqr spectrum. The radial velocity curve in figure~\ref{fig:rvcurve} was then obtained by fitting a sinusoid through the CCF peaks using the equation
\begin{equation}
V(\phi) = \gamma + K_{r}\sin{2\pi\phi}
\end{equation}
\noindent where $V(\phi)$ is the radial velocity of the secondary at phase $\phi$, $\gamma$ the systemic velocity and $K_{r}$ the semi-amplitude of the secondary star.

From this analysis a new zero-point was obtained for the ephemeris of
\begin{equation}
\mathrm{T}_{0} = \mathrm{HJD} 2452131.31617165\pm0.000092345
\end{equation}
\noindent with the orbital period fixed at $\mathrm{P} = 0.41165553 \mathrm{d}$ (from \citealt{casares1996}). 

All data used in this work has been phased with respect to this new ephemeris. While the cross-correlation method is relatively insensitive to the use of an ill-matching template or incorrect amount of broadening, the effects of irradiation are more likely to dominate measurements, introducing systematic errors in radial velocity if not accounted for (e.g. \citealt{davey1992}). The ephemeris derived here provides a substantially improved image quality (both in terms of final reduced $\chi^{2}$ and artefacts in the reconstructed map) when compared against that published by \cite{watson2006}. 

For the purposes of this work we have not attempted a full spectral-type and binary determination, however, for completeness we obtained a secondary star radial velocity amplitude of $\mathrm{K_{r}} = 167.433\pm0.022 \mathrm{kms^{-1}}$, a systemic velocity of $\gamma = -62.07\pm0.02 \mathrm{kms^{-1}}$ and a mean $v\sin{i}$ from the 4 nights observations of $101.5\pm2.0$ kms\tsu{-1}. This assumes the K4V template star HD 187760 has a systemic velocity of $-21.545\pm0.003 \mathrm{kms^{-1}}$, as measured by a Gaussian fit to the LSD line profile of the K4V template star (using a line-list, where lines with a central depth shallower than 10 per cent of the continuum were excluded).

\begin{figure}
\centering
\includegraphics[width=0.5\textwidth]{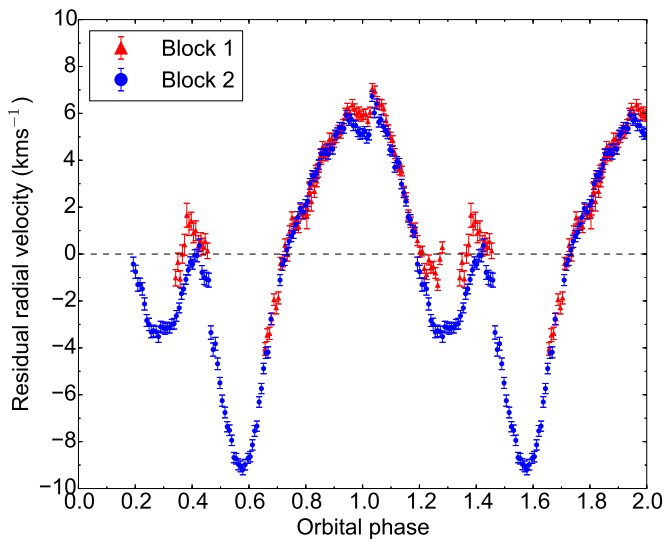}
\caption{Residuals (phase-folded and repeated) after subtraction of the fitted sinusoid to the RV curve in figure~\ref{fig:rvcurve}. Red triangles are block 1 data, blue circles are block 2 data taken 9 days later. RV residuals can be traced to individual features in the Roche tomograms presented later, and differences in the residuals between blocks 1 and 2 are attributed to both spot evolution and differential rotation.}
\label{fig:rvcurveresiduals}
\end{figure}

Figure~\ref{fig:rvcurve} shows the measured radial velocity for each data block as well as the fitted sinusoid. The velocities are noticeably lower around phase 0.25 due to a systemic velocity offset between blocks 1 and 2 (see section~\ref{sec:incsystemicmasses}). This conventional approach to binary parameter measurements are inherently systematically biased due to surface inhomogeneities (such as irradiation and starspots) and tidal distortion, which causes the centre-of-light (COL) to no longer coincide with the centre-of-mass (COM). This inherent bias is clearly shown in the deviations from a perfect sinusoid in figure~\ref{fig:rvcurve}, and by examining the residuals after subtracting the fitted sinusoid (see figure~\ref{fig:rvcurveresiduals}), surface features can be identified. The peak around phase 0.4 and the trough around phase 0.6 correspond to a region of reduced absorption-line strength on the inner hemisphere of the secondary, and is most likely caused by irradiation from the accretion regions in addition to starspots, both moving the COL away from the COM. The large peak around phase 1 and the trough around phase 1.3 indicate a region of enhanced absorption-line strength on the outer hemisphere, again moving the COL away from the COM, and could be caused by very large starspots. The scatter in RVs between block 1 and block 2 may indicate to spot evolution or differential rotation, and is an independent indication of surface features from Roche tomography (see section~\ref{sec:surfacemaps}). Due to the bias in these measurements, the parameters derived from the radial velocity curve have not been used in the subsequent analysis presented in this work.

\subsection{Continuum fitting}
\label{sec:contfit}
In order to generate the least-squares deconvolution line profiles required for Roche tomography, the continuum must be flattened. In order to achieve this we fit the continuum. For this and all subsequent analysis, emission lines, tellurics and regions considered too noisy were masked in the spectra (note emission lines are not required for Roche tomography since we are mapping the secondary star, not the disk). The masked spectral regions are shown in table~\ref{tab:wavemasked}.

\begin{table}
\centering
\caption{Spectral regions not included in continuum fitting and LSD process.} 
\label{tab:wavemasked}
\begin{tabular}{lllllll}
\toprule
Colour & Masked region (\r{A})	& Comments \\
\midrule
Blue			&	$< 3990$		&	Noise, emission lines\\
			&	$4076 - 4130$	&	H delta emission\\
			&	$4312 - 4372$	&	H gamma emission\\
			&	$> 4517$		&	No flux\\
Red lower		&	$<4789$		&	No flux\\
			&	$4830 - 4885$	&	H beta emission\\
			&	$5006 - 5022$	&	He I emission\\
			&	$>5762$		&	No flux\\
Red upper		&	$<5910$		&No flux, \\
				&				&He I emission \\
				&				&Na I doublet\\
			&	$6270 - 6320$	&	Tellurics\\
			&	$6520 - 6620$	&	H alpha emission\\
			&	$6640 - 6720$	&	He I emission\\
			&	$> 6786$		&	No flux\\
\bottomrule
\end{tabular}
\end{table}

The contribution of the secondary star to the total system luminosity is constantly varying, and there is an unknown contribution to each spectrum due to variable light from the accretion regions. A master continuum fit to the data (such as that done by \citealt{cameron1994}) is not appropriate due to the constantly changing continuum slope, caused by (for example) flaring, or from the varying aspect of the accretion region. Additionally, normalisation of the continuum would result in the photospheric absorption lines from the secondary varying in relative strength between exposures. This forces us to subtract the continuum from each spectrum. 

Slit loss corrections could be undertaken by placing another star in the slit, but unfortunately this is not possible here due to the cross-dispersed nature of the echelle spectrograph. For this reason, when conducting Roche tomography it is normal to obtain photometry to monitor transparency and target brightness variations. Unfortunately for these observations we were unable to obtain simultaneous photometry, and so another means of slit correction needed to be applied. For this we employed a method of relative scaling of adjacent LSD profiles, which is discussed in detail in section~\ref{sec:scaling}.

Since we cannot construct a master continuum fit, an algorithm was developed to place spline points at regular intervals in the spectra. During this process, emission lines from the accretion disk and tellurics were masked out (see table~\ref{tab:wavemasked} for details of excluded regions).

It is difficult to determine where the true continuum lies due to the forest of broadened and blended lines which act to lower the apparent continuum level. We therefore adopted an algorithm which tries to identify regions of line-free continuum. In order to do this we adopted the following strategy.

\begin{itemize}
\item The spectra were first split into discrete wavelength windows spanning 40 \AA~.
\item A first order polynomial was fit to each window individually and all the data points below the fit were rejected.
\item Another first order polynomial was fit to the remaining data points in each separate window.
\item The highest flux value within $3\sigma$ of this latter fit was determined, and a constant was added to the fit so that it now passed through this flux value.
\item A spline point was then positioned on the linear fit at the central wavelength of the window in question.
\end{itemize}

We found that deep absorption lines could systematically lower the fit since a large number of data points would lie below the true continuum level, fooling the above procedure into placing a spline point too low. To correct for this, spline points placed in deep lines were identified and rejected by comparing it to the points on either side and to the previously included spline point. Spline points were placed on either side of masked regions, and the spline fit was allowed to vary freely in this region.

The results of this fitting process were tested and checked visually until an acceptable fit was acquired. This is obviously very subjective as it is difficult to judge where the true continuum actually lies when dealing with heavily broadened and blended lines, which effectively acts to suppress the apparent continuum level. The normalised spectra were compared visually in sequential order, and the continuum fit was found to be stable.

The impact of a poorly-fitted continuum was previously assessed by \cite{watson2006}, who found that even continuum fits which were obviously incorrect did not adversely affect the shape of the line profiles after carrying out the least squares deconvolution (LSD) process described in section~\ref{sec:lsd}. However, we found that the continuum was systematically fit at too high a level, leading to LSD profiles with continuum regions lying significantly below zero (as also found by \citealt{watson2006}). This was solved by shifting the continuum fit to a lower level until the LSD profiles lay at zero. The line shape was not affected by this process. 

\section{Roche tomography}
\label{sec:rochetomography}
Roche tomography is analogous to Doppler imaging (e.g. \citealt{vogt1983}) and has been successfully applied to the donor stars of CVs over the past 20 years (\citealt{rutten1994}, \citealt{rutten1996}, \citealt{watson2003}, \citealt{schwope2004}, \citealt{watson2006}, \citealt{watson2007}, \citealt{dunford2012}). 

Roche tomography is specifically designed to map Roche-lobe-filling secondary stars in interacting binaries such as CVs and X-ray binaries (e.g. \cite{shahbaz2014}, and assumes the secondary is locked in synchronous rotation and has a circularised orbit. Rather than repeat a detailed description of the methodology and axioms of Roche tomography here, we refer the reader to the references above and the technical reviews of Roche tomography by \citet{watson2001} and \citet{dhillon2001}. 

Relevant points of note with respect to this work are that we select a unique map by employing the maximum-entropy MEMSYS algorithm developed by \cite{skilling1984}. We use a moving uniform default map, where each element is set to the mean value of the reconstructed map. We do not adopt a two-temperature or filling-factor model (e.g. \citealt{cameron1994}), since donor stars are expected to exhibit large temperature differences due to irradiation by the primary. The lack of a two-temperature model means our Roche tomograms may be prone to the growth of bright pixels which make the tomograms quantitatively more difficult to analyse.

\section{Least squares deconvolution}
\label{sec:lsd}
Spot features appear in line profiles as an emission bump (actually a lack of absorption), and are typically a few percent of the line depth. This means that very high signal-to-noise data are required, which cannot directly be achieved for a single line with AE Aqr due to its faintness and the requirements for short exposures to avoid orbital smearing. Thus, the technique of Least Squares Deconvolution (LSD) was employed which effectively stacks the $\sim1000$'s of photospheric absorption lines observable in a single echelle spectrum to produce a `mean' profile of greatly increased signal-to-noise. Theoretically, the multiplex gain in signal-to-noise is the square root of the number of lines observed. The technique is well documented and was first applied by \citet{donati1997} and has since been used in many Doppler imaging studies (e.g. \citealt{jeffers2002, barnes2004, marsden2005}) as well as in the mapping of starspots on CVs (e.g. \citealt{watson2007, dunford2012}). For more details on LSD, see these references as well as the review by \citet{cameron2001}.

The technique assumes all photospheric lines have the same local line profile shape, and that starspots affect all of the rotationally broadened line profiles in the same way, so the morphology of the characteristic starspot bump is identical. The positions and relative strengths of observed lines in each echelle spectrum must be known. For this work, a line list generated by the Vienna Atomic Line Database (VALD) was used (see \citealt{kupka1999, kupka2000}). The spectral type of AE Aqr has been determined to lie in the range K3--K5 V (\citealt{crawford1956, chincarini1981, tanzi1981, bruch1991}), and so a line-list for a stellar atmosphere with $\mathrm{T_{eff}} = 4750 \mathrm{K}$ and $\mathrm{log}g = 4.5$ (the closest approximation available to a K4V spectral type) was obtained and used in the LSD process. 

Initially the LSD line profiles exhibited a slope in the continuum. After trials it was found that including more lines in the deconvolution acted to reduce the slope in the continuum region of the deconvolved line profiles. A detection limit of 0.2 (of the normalised line depth, below which all lines with a smaller central depth were excluded) was adopted, giving 3202 lines over which to carry out LSD. Including more lines resulted in minimal improvement but significantly increased computation time. Since the line-list obtained from VALD contains normalised line-depths, where as Roche tomography uses continuum subtracted spectra, each line-depth was scaled by a fit to the continuum of a K4V template star, meaning each line's relative depth was correct for use with continuum subtracted data.

The exact choice of line-list is unlikely to affect the results presented here as \citet{barnes1999} found that the LSD process was insensitive to the use of an incorrect line-list (see \citealt{watson2006} for more detail).

After this process, there was still evidence of a continuum slope in the LSD profiles. This was removed by masking out the centre of the line and subtracting a second-order polynomial which was fit to the continuum. This smooth function did not significantly alter the profiles' shape. 

\subsection{Scaling of profile depths}
\label{sec:scaling}
As previously mentioned, Roche tomography cannot be performed on data which has not been corrected for slit losses. Normally, when conducting Roche tomography, simultaneous photometry is taken to monitor transparency and target brightness variations, and due to its absence here, the LSD line profiles were scaled relative to each other using an adjacent-profile scaling approach. For this we assumed that the line profile does not change significantly from profile to profile, apart from small changes due to rotational effects. We also assume that any large changes are due to slit losses and transparency variations, rather than changing spot features.

For the line profile scaling we employed an optimal subtraction method, where one profile is scaled and subtracted from its neighbouring profile, the optimal scaling factor being that which minimises the residuals. This scaling factor was found between a benchmark profile (chosen for its near-Gaussian shape) and its neighbour. The factor was applied to the neighbouring LSD profile and the process repeated for the entire time series of LSD profiles, using the newly scaled profile as the comparison for the next. It was found that applying a slight Gaussian smoothing to the line profile allowed for better scaling as this reduced the effect of noise and of a changing line profile shape. The reader should be reminded that Roche tomography uses relative fluxes, and so only the shape, not the absolute depth of the line profiles matter.

The above scaling technique was extensively tested in simulations. For this we constructed a synthetic dataset which resembled the final AE Aqr tomogram. We took each line profile of our scaled dataset (from above) and scaled them by varying amounts to simulate slit losses and transparency variations. We then applied the adjacent-profile scaling method and compared our calculated scaling factors to those that were injected at the start of the simulation. 

The initial adjacent-profile scaling was typically correct to within 4 per cent for both the simulated block 1 and block 2 datasets, with a particular worst case scenario, discussed below, that leads to a 16 per cent error. This occurs around phase 0.5 and is due to rapidly changing line depths caused by the irradiation around the L\tsb{1} point. These poorly scaled profiles showed up in the Roche tomograms as prominent artefacts in the the form of two large dark spots on the southern hemisphere. The systematic errors in this method were corrected as follows

\begin{itemize}
\item A map was reconstructed from the data.
\item Random scaling factors were applied to the computed fits to the line profiles.
\item The adjacent-profile scaling method above was applied to these line profiles (starting from a benchmark profile).
\item The scaling factors between the computed profiles and those from the previous step were found.
\item These scaling factors were divided by the random scaling factor assigned to the benchmark profile, giving the percentage to which they were correct.
\item Then, the scaling factors between the observed line profiles and the reconstructed line profiles were found.
\item These latter scaling factors were divided by the percentage above, giving the correction factors that need to be applied after using the adjacent-profile scaling method.
\end{itemize}

\noindent The systematic corrections were found to be typically 5 per cent, and 13 per cent in the worst case. The resulting profiles were used in the reconstruction process in Roche tomography. The computed line profiles were compared with the input profiles, and we found profiles around phase 0.5 were poorly fitted. We then scaled the input profiles to match the computed profiles. We note that, even including the poorly scaled profiles, the spot features in the northern hemisphere are not significantly altered.

\begin{figure*}
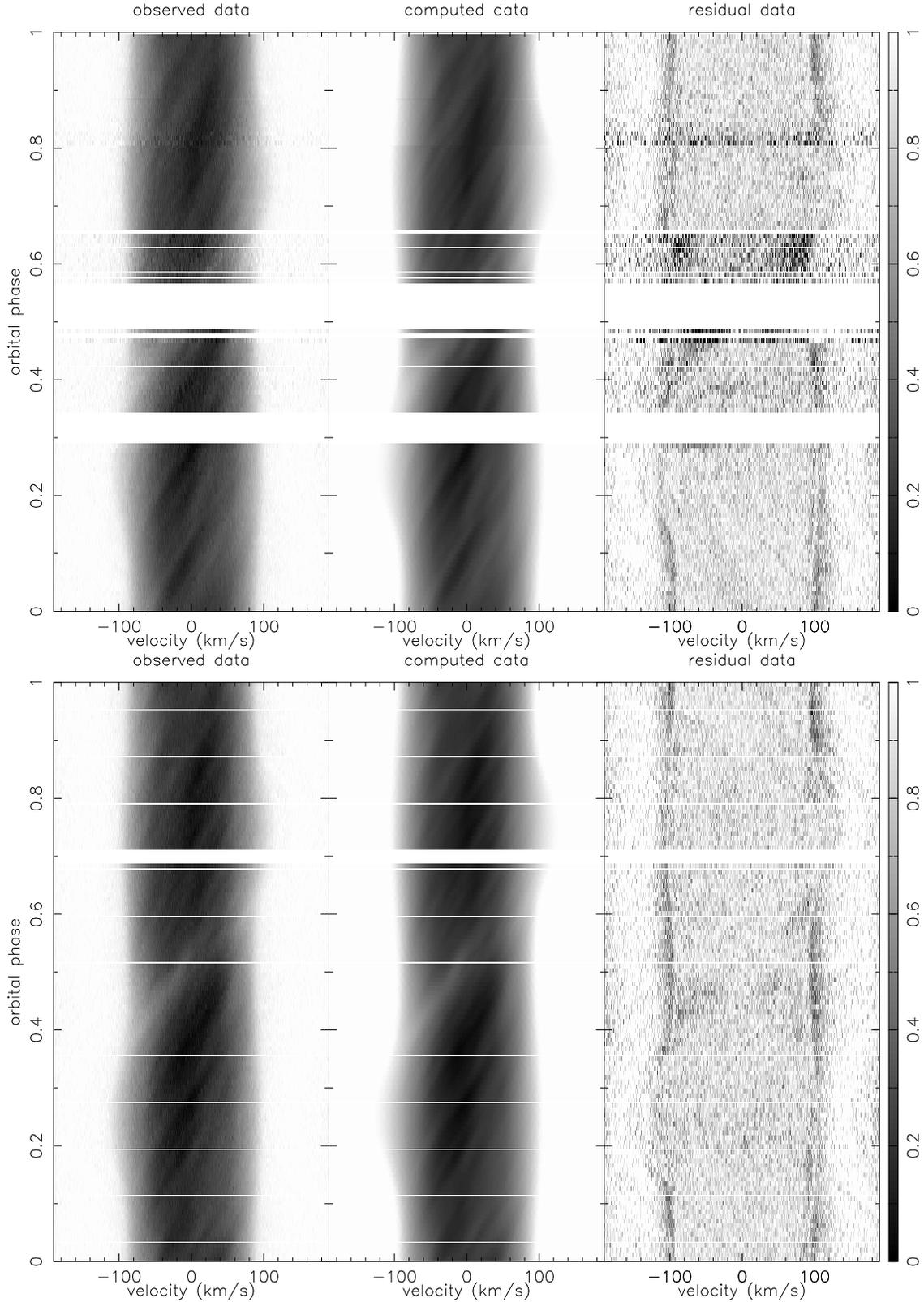

\centering
\includegraphics[width=0.6\textwidth,angle=270]{n1n2_trails.ps}
\includegraphics[width=0.6\textwidth,angle=270]{n3n4_trails.ps}
\caption{Trailed spectra of AE Aqr (top panel is block 1 data, bottom panel is block 2 data). Panels show (from left to right) the observed LSD data, computed data from the Roche tomography reconstruction and the residuals (increased by a factor of 10). Starspot features in these panels appear bright. A grey-scale wedge is also shown, where a value of 1 corresponds to the maximum line depth in the reconstructed profiles. The orbital motion has been removed assuming the binary parameters found in section~\ref{sec:systempars}, which allows the individual starspot tracks across the profiles and the variation in $V_{rot}\sin{i}$ to be more clearly observed.}
\label{fig:trails}
\end{figure*}

The final LSD profiles are trailed in figure~\ref{fig:trails}, showing the emission bumps due to starspots moving from blue (negative velocities) to red (positive velocities) through the profile as AE Aqr rotates, in addition to the secondary stars orbital motion and variations in the projected equatorial velocity, $v\sin{i}$.

The residuals in figure~\ref{fig:trails} show narrow emission features in the form of sinusoids across the entire orbitally-corrected trail for both observation blocks. These appear as vertical lines when the orbital motion has not been removed, and the most prominent of these features can be visually tracked through sequential line profiles. These emission bumps are also seen lying outside the stellar absorption profile, off the stellar limb, and therefore cannot lie on the stellar surface. We suggest that the most likely cause for these features is the presence of circumstellar material in the form of prominences. A similar feature was also observed in the LSD line profiles of BV Cen, with the conclusion that this was due to a slingshot prominence (see \citealt{watson2007} for further discussion). The emission features seen here, however, are too faint to carry out any further conclusive analysis.

\section{System parameters}
\label{sec:systempars}
Adoption of incorrect system parameters such as systemic velocity, component masses and orbital inclination when carrying out a Roche tomography reconstruction results in spurious artefacts in the final map (see \citealt{watson2001} for details). Artefacts normally appear as bright and dark streaks, increasing the amount of structure mapped in the final image, which in turn leads to a decrease in the entropy regularisation statistic.

We can constrain the binary parameters by carrying out reconstructions for many pairs of component masses (iterating to the same $\chi^{2}$), creating an `entropy landscape' (see figures~\ref{fig:entlandblock1}~\&~\ref{fig:entlandblock2}) where each grid tile corresponds to the map entropy value for the reconstruction for that pair of component masses. Entropy landscapes can also be repeated for different values for the orbital inclination $i$ and the systemic velocity $\gamma$. One then adopts the set of parameters ($M_1, M_2, i, \gamma$) that produce the map containing the least structure (the map of maximum entropy).

\subsection{Limb Darkening}
\label{sec:limbdarkening}
After trial reconstructions, it was found that the square-root limb darkening law as used in previous work (see \citealt{watson2006}) resulted in a poor fit to the wings of the observed LSD profiles. The mismatch to the wings is evident in this dataset due to the superior signal-to-noise compared to previous Roche tomography datasets.

For this work we adopted a four-parameter non-linear model (see \citealt{claret2000}), given by

\begin{align}
\frac{I(\mu)}{I(1)} &= 1 - a_{1}(1-\mu^\frac{1}{2}) - a_{2}(1-\mu) - a_{3}(1-\mu^\frac{3}{2}) - a_{4}(1-\mu^{2})
\label{eq:limbdarkeninglaw}
\end{align}

\noindent where $\mu = \cos{\gamma}$ ($\gamma$ is the angle between the line of sight and the emergent flux), and $I(1)$ is the monochromatic specific intensity at the centre of the stellar disk. 

In order to calculate the correct limb darkening coefficients we determined the effective central wavelength of the UVES data using,

\begin{equation}
\lambda_{\mathrm{cen}} = \frac{\sum_{i}\nolimits \frac{1}{\sigma_{i}}\:d_{i}\:\lambda_{i}}{\sum_{i}\nolimits\:\frac{1}{\sigma_{i}}\:d_{i}}
\label{eq:cenwave}
\end{equation}

\noindent where $d_{i}$ is the line depth at wavelength $\lambda_{i}$, and $\sigma_{i}$ the error in the data at~$\lambda_{i}$.

Since there are effectively three wavelength ranges in the UVES setup (the blue CCD and the two halves of the red CCD, red-upper and red-lower) we calculated $\lambda_{cen}$ for all the exposures separately, for each CCD. We then averaged these for each CCD, and averaged the resulting three values, giving the final central wavelengths as $\lambda_{cen} = 5474.206$~\AA~ and $\lambda_{cen} = 5503.454$~\AA~ for block 1 and block 2, respectively.

Once $\lambda_{cen}$ was found, the limb darkening coefficients were then determined by linearly interpolating between the tabulated wavelengths given in \cite{claret2000}. We adopted stellar parameters closest to that of a K4V star, which for the PHOENIX model coefficients were $\log{g} = 4.5$ and $T_{eff} = 4800 \mathrm{K}$. We also trialled limb darkening coefficients from the ATLAS model, but found that maps reconstructed using the PHOENIX coefficients resulted in reconstructions to higher entropy values (by around 10\%). We therefore adopted the PHOENIX coefficients for the rest of the work, which are listed in table~\ref{tab:limbcoeff} for both the separate observing blocks.

\begin{table}
\centering
\caption{Limb-darkening coefficients used in equation~\eqref{eq:limbdarkeninglaw}.}
\label{tab:limbcoeff}
\begin{tabular}{crr}
\toprule
Coefficient & Block 1 & Block 2 \\
\midrule
a1	&	0.7253	&	0.7254 \\
a2	&	-0.7380	&	-0.7348 \\
a3	&	1.3314	&	1.3288 \\
a4	&	-0.4264	&	-0.4281 \\
\bottomrule
\end{tabular}
\end{table}

\subsection{Systemic velocity, inclination and masses}
\label{sec:incsystemicmasses}
The binary parameters were found independently for each observation block. A sequence of entropy landscapes were constructed for a range of orbital inclinations \emph{i} and systemic velocities $\gamma$. For each $i$ and $\gamma$ we chose the map of maximum entropy in the corresponding entropy landscape. We present the results of the entropy landscape analyses below.

\subsubsection{Systemic velocity}
\begin{figure}
\centering
\includegraphics[width=0.5\textwidth]{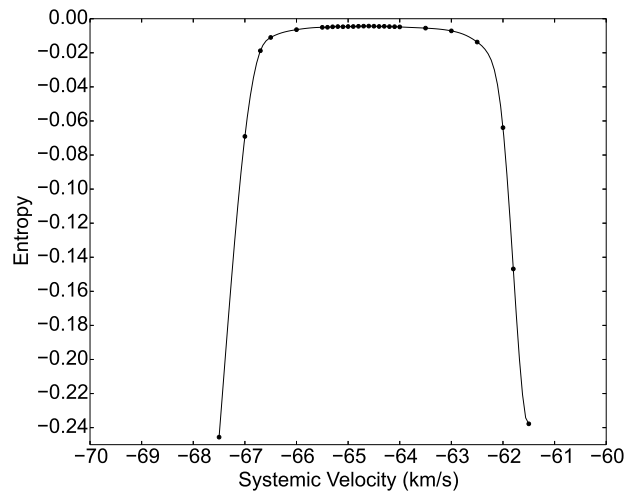}
\caption{Points show the maximum entropy value obtained in each entropy landscape for block 1 data as a function of systemic velocity, assuming an orbital inclination of $50^{\circ}$. The solid line shows the general trend, with a maximum of $\gamma_{\mathrm{block\:1}} = -64.7\pm 2.1$ kms\tsu{-1}.}
\label{fig:block1systemic}
\end{figure}

\begin{figure}
\centering
\includegraphics[width=0.5\textwidth]{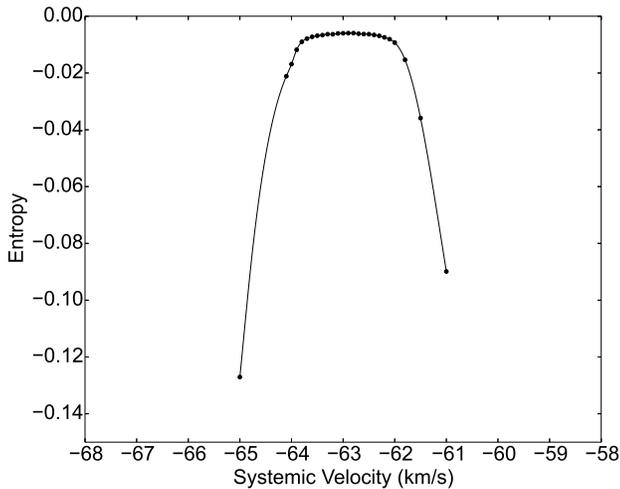}
\caption{As figure~\ref{fig:block1systemic}, but for block 2, assuming an orbital inclination of $51^{\circ}$. The solid curve shows the general trend, with a maximum of $\gamma_{\mathrm{block\:2}} = -62.9\pm 1.0$ kms\tsu{-1}.}
\label{fig:block2systemic}
\end{figure}

Figures~\ref{fig:block1systemic}~\&~\ref{fig:block2systemic} show the systemic velocities yielded; $\gamma_{\mathrm{block\:1}} = -64.7\pm 2.1$ kms\tsu{-1} and $\gamma_{\mathrm{block\:2}} = -62.9\pm 1.0$ kms\tsu{-1}, where the error bars are estimated from the spread of velocities which produce maps of similar entropy (the "entropy plateau" seen in figures~\ref{fig:block1systemic}~\&~\ref{fig:block2systemic}). These measurements are consistent with that of previous work (see table~\ref{tab:systemparameters}). The difference between the two blocks of 1.8 kms\tsu{-1} can be explained by an instrumental offset, since offsets of $\sim100$ms\tsu{-1} are frequently reported for wavelength stabilised high-precision echelle spectra such as SOPHIE (see \citealt{simpson2011}).

The value obtained using Roche Tomography is similar to that obtained from the radial velocity curve of $\gamma_{\mathrm{rv}} = -62.07\pm0.02 \mathrm{kms^{-1}}$, although the RV curve measurement will be biased (see section~\ref{sec:ephemeris}). The value obtained for the systemic velocity from the entropy landscapes is independent of the assumed inclination, as has been found previously by \cite{watson2003, watson2006}.

\begin{table*}
\caption{System parameters as found by the respective authors.}
\label{tab:systemparameters}
\begin{tabular}{cccccc}
\toprule
Author & Systemic velocity $\gamma$ (kms\tsu{-1}) & Inclination $i$ (degrees) & $M_{1}$ (M$_{\odot}$) & $M_{2}$ (M$_{\odot}$) & Mass ratio $q = \sfrac{\mathrm{M}_{2}}{\mathrm{M}_{1}}$\\
\midrule
Block 1 (this work) 	&	$-64.7\pm 2.1$&	50		&1.20	&0.81	&0.68\\
Block 2 (this work)	&	$-62.9\pm 1.0$&	51		&1.17	&0.78	&0.67\\
\cite{echevarria2008}&	$-63$		&	70		&0.63	&0.37	&0.60 \\	
\cite{watson2006}	&	$-63$		&	66		&0.74	&0.50	&0.68\\
\cite{casares1996}	&	$-60.9\pm 2.4$&	$58\pm6$&$0.79\pm0.16$&$0.50\pm0.10$&0.63\\
\cite{welsh1995}	&	$-63\pm 3$	&	$54.9\pm7.2$&$0.89\pm0.23$&$0.57\pm0.15$&0.64\\
\bottomrule
\end{tabular}
\end{table*}

\subsubsection{Inclination}
\begin{figure}
\centering
\includegraphics[width=0.5\textwidth]{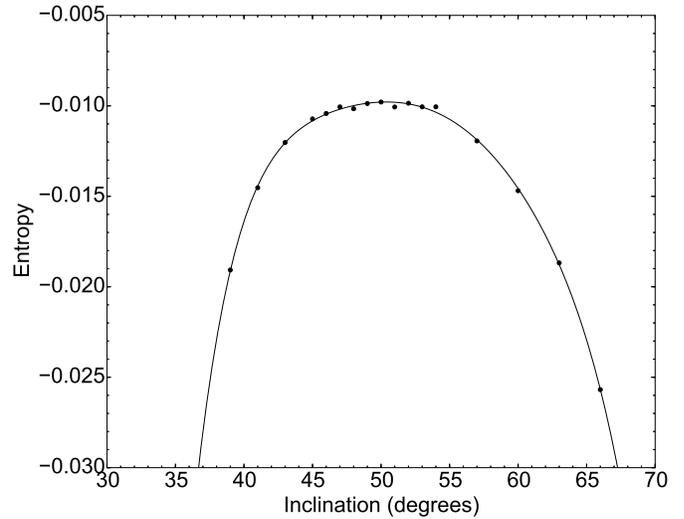}
\caption{Points show the maximum entropy value obtained in each entropy landscape as a function of inclination for block 1 data, assuming a systemic velocity of $\gamma_{\mathrm{block\:1}} = -64.7$ kms\tsu{-1}. The solid curve shows a sixth order polynomial fit through these points, as a guild only to highlight the maximum of $i_{\mathrm{block\:1}} = 50^{\circ}$.}
\label{fig:block1inclination}
\end{figure}

\begin{figure}
\centering
\includegraphics[width=0.5\textwidth]{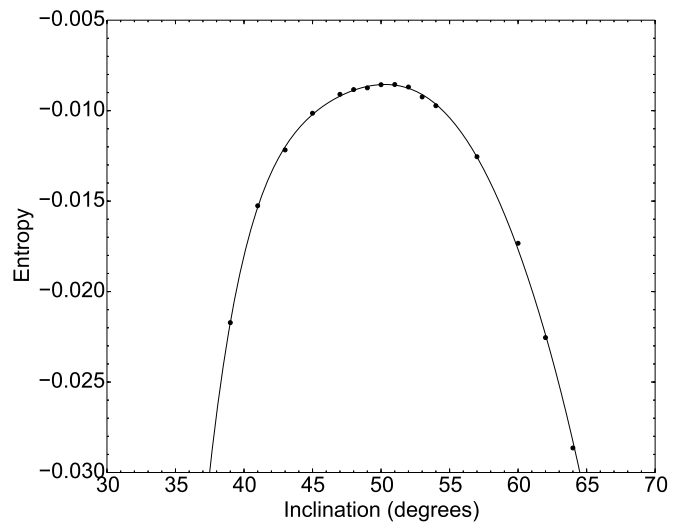}
\caption{The same as figure~\ref{fig:block1inclination}, but for block 2 data, assuming a systemic velocity of $\gamma_{\mathrm{block\:2}} = -62.9$ kms\tsu{-1}. The solid curve shows a sixth order polynomial fit through these points, highlighting the maximum of $i_{\mathrm{block\:2}} = 51^{\circ}$.}
\label{fig:block2inclination}
\end{figure}

Figures~\ref{fig:block1inclination}~\&~\ref{fig:block2inclination} show the maximum entropy value obtained as a function of inclination from the entropy landscape analyses, assuming systemic velocities of $\gamma_{\mathrm{block\:1}} = -64.7$ kms\tsu{-1} and $\gamma_{\mathrm{block\:2}} = -62.9$ kms\tsu{-1} as derived above for block 1 and block 2, respectively. We obtain inclinations of $i = 50^\circ$ for block 1 and $i = 51^\circ$ for block 2.

The inclination measurements found here are lower than previous estimates, but still in agreement with those found by \cite{welsh1995} and \cite{casares1996} (see table~\ref{tab:systemparameters}).

Work by \cite{echevarria2008} suggests a higher inclination up to $70^{\circ}$ but, perhaps more importantly, a previous Roche tomography study of AE Aqr by \cite{watson2006} yielded a higher  inclination of $66^{\circ}$, despite using the same methods and approaches as this work.

It should be noted, however, that the inclination is the worst constrained parameter when using Roche tomography, but it is reassuring that the two independent data sets agree to within $1^{\circ}$ on the inclination of $50^{\circ}$ to $51^{\circ}$, and that this lies below the $i < 70^{\circ}$ limit inferred from the lack of eclipses (\citealt{chanan1976}). In addition, the reconstructed maps (see section~\ref{sec:surfacemaps}) at both $50^{\circ}$ and $51^{\circ}$ and at the previously determined value of $66^{\circ}$ show very similar features, and the calculated differential rotation rate (see section~\ref{sec:diffrot}) is similar. This shows that the surface features mapped by Roche tomography, as well as the inferred differential rotation rate calculated later, are robust even when the orbital inclination is varied over the full span of previously reported values.

\subsubsection{Masses}
\begin{figure}
\centering
\includegraphics[width=0.4\textwidth,angle=270]{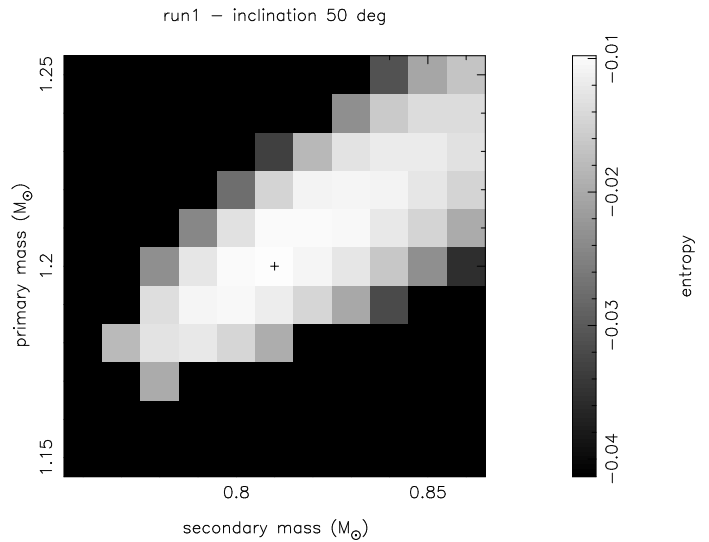}
\caption{The entropy landscape for AE Aqr using data from the first observation block, assuming an orbital inclination of $i = 50^{\circ}$ and a systemic velocity of $\gamma_{\mathrm{block\:1}} = -64.7$ kms\tsu{-1}. Dark regions indicate masses for which no acceptable solution could be found. The cross marks the point of maximum entropy, corresponding to component masses of $\mathrm{M}_1 = 1.20 \mathrm{M}_{\odot}$ and $\mathrm{M}_2 = 0.81 \mathrm{M}_{\odot}$.}
\label{fig:entlandblock1}
\end{figure}

\begin{figure}
\centering
\includegraphics[width=0.4\textwidth,angle=270]{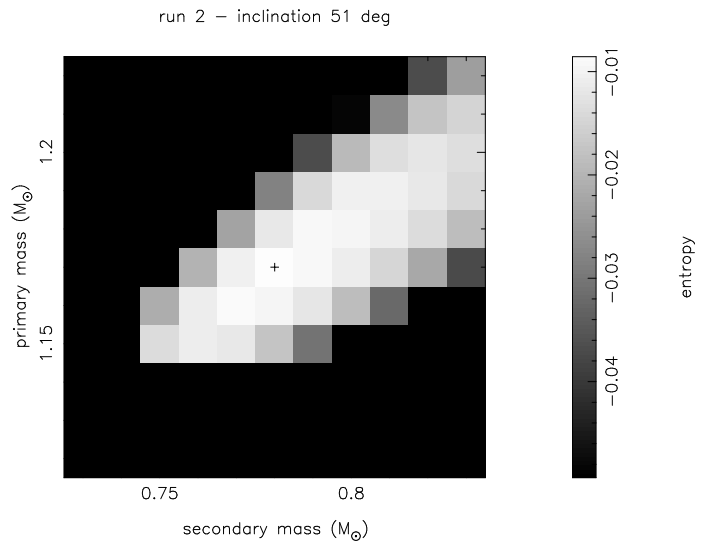}
\caption{The same as figure~\ref{fig:entlandblock1}, but this time showing the entropy landscape for AE Aqr using data from the second observation block, assuming an orbital inclination of $i = 51^{\circ}$ and a systemic velocity of $\gamma_{\mathrm{block\:2}} = -62.9$ kms\tsu{-1}. The point of maximum entropy corresponds to component masses of $\mathrm{M}_1 = 1.17 \mathrm{M}_{\odot}$ and $\mathrm{M}_2 = 0.78 \mathrm{M}_{\odot}$.}
\label{fig:entlandblock2}
\end{figure}

The entropy landscapes for AE Aqr are shown in figures~\ref{fig:entlandblock1}~\&~\ref{fig:entlandblock2} for block 1 and block 2, respectively. For block 1 we assume $i = 50^{\circ}$ and $\gamma_{\mathrm{block\:1}} = -64.7$ kms\tsu{-1}, and from this we derive a secondary mass of 0.81 M$_{\odot}$ and a primary mass of 1.20 M$_{\odot}$. For block 2 we assume $i = 51^{\circ}$ and $\gamma_{\mathrm{block\:2}} = -62.9$ kms\tsu{-1}, and from this we derive a secondary mass of 0.78 M$_{\odot}$ and a primary mass of 1.17 M$_{\odot}$. These masses are consistent within this work, but are much larger when compared with previous estimates (see table~\ref{tab:systemparameters}), due to the inclination difference. It should be noted, however, that the mass ratios derived here are in good agreement with previous work.

\begin{figure}
\centering
\includegraphics[width=0.4\textwidth,angle=0]{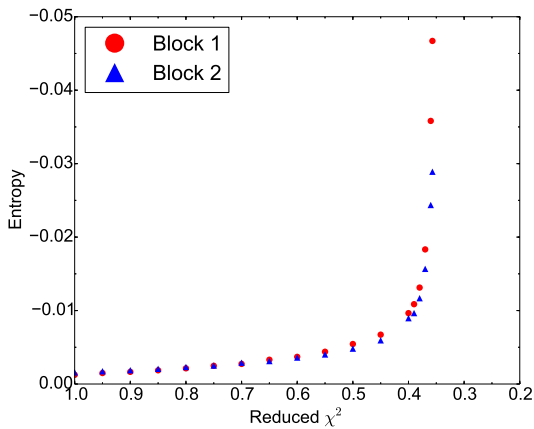}
\caption{Reconstructed-map entropy as a function of reduced $\chi^{2}$, for both observation blocks, using the system parameters derived in section~\ref{sec:incsystemicmasses}. Note the sharp decrease in entropy below a reduced $\chi^{2}$ of 0.4.}
\label{fig:chivsentropy}
\end{figure}

In all reconstructions we fit to a reduced $\chi^{2} = 0.4$, which indicates that our propagated error bars are systematically overestimated for the scaled profiles, but this does not affect the reconstructions. This aim $\chi^{2}$ was chosen as the entropy of the reconstructed maps dramatically decreases when fits to $\chi^{2}$ lower than this are performed (see figure~\ref{fig:chivsentropy}). This is due to a marked increase in the presence of small-scale structure due to the mapping of noise features in the Roche tomogram. On the other hand, fitting to a higher reduced $\chi^{2}$ causes the reconstructed maps to have less distinct structure, and the inclination and masses are not constrained as more pixels are assigned the default map value.

Assigning errors to the derived system parameters obtained from the entropy landscapes is not straight forward. As discussed in \cite{watson2006} and \cite{watson2001}, it would require using a Monte-Carlo style technique combined with a bootstrap resampling method (\citealt{efron1979, efron1993}) to generate synthetic datasets drawn from the same parent population as the observed dataset. Then the same analysis carried out in this work would need to be applied to the hundreds of bootstrapped datasets, requiring several months of computation. This is unfeasible and so we do not assign strict error bars to our derived system parameters.

\section{Surface maps}
\label{sec:surfacemaps}

\subsection{Global features}
\begin{figure}
\centering
\includegraphics[width=0.5\textwidth]{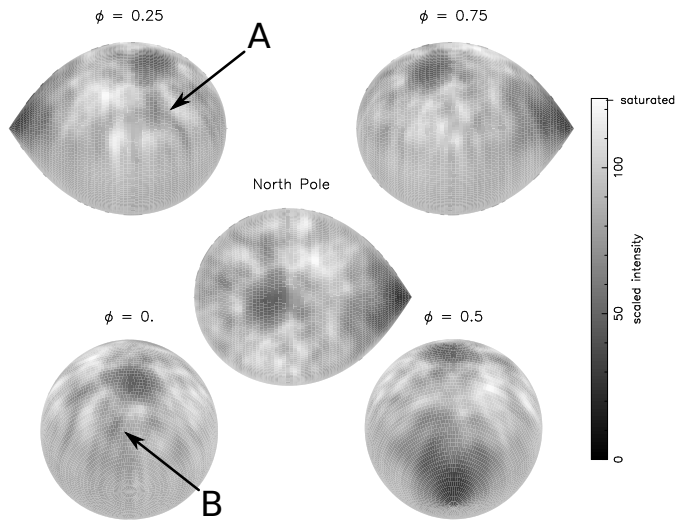}
\caption{The Roche tomogram of AE Aqr using block 1 data. Dark grey scales indicate regions of reduced absorption line strength that is due to either the presence of starspots or the impact of irradiation. The orbital phase is indicated above each panel. Roche tomograms are shown without limb darkening for clarity.}
\label{fig:block1tomogram}
\end{figure}

\begin{figure}
\centering
\includegraphics[angle=270,width=0.5\textwidth]{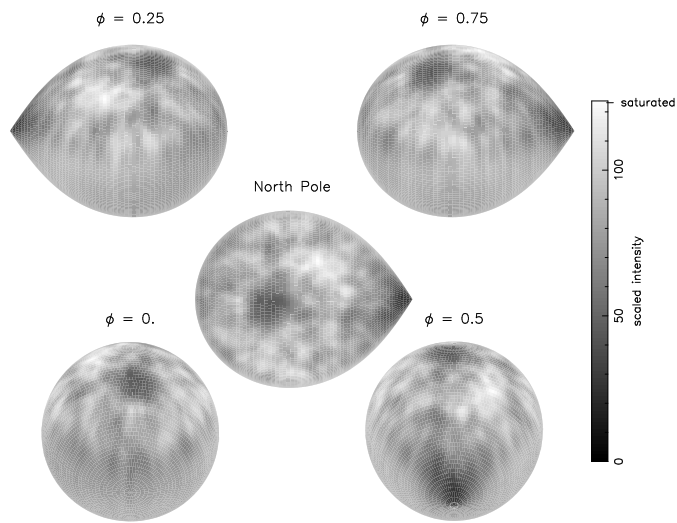}
\caption{The same as figure~\ref{fig:block1tomogram} but displaying the Roche tomogram of AE Aqr reconstructed using the data obtained on observing block 2 (9 days after observing block 1).}
\label{fig:block2tomogram}
\end{figure}

Using the system parameters derived in section~\ref{sec:systempars} we have constructed Roche tomograms of the secondary star in AE Aqr for both observation block 1 and block 2 of our dataset (see figures~\ref{fig:block1tomogram}~\&~\ref{fig:block2tomogram}). The corresponding fit to the datasets obtained on both nights are displayed in figure~\ref{fig:trails}.

Dark spots features are evident in both tomograms. Common to both maps is the large, high-latitude starspot located above $60^{\circ}$ latitude. Similar high-latitude spots are commonly found in Doppler imaging studies of rapidly-rotating single stars such as AB Dor (\citealt{hussain2007}), LQ Hya (\citealt{donati1999}) and PZ Tel (\citealt{barnes2000}), and have also been found in Roche tomograms of other CVs including RU Peg (\citealt{dunford2012}), BV Cen (\citealt{watson2007}) and a previous study of AE Aqr (\citealt{watson2006}). An explanation for these high-latitude spots is given by \cite{schuessler1992} who suggest that strong Coriolis forces in these rapidly rotating stars drive the magnetic flux tubes towards the polar regions. Alternatively, spots may migrate polewards after forming at lower latitudes, as has been found for the RS CVn binary HR1009 (\citealt{vogt1999}).

We note that in all previous Roche tomograms of CV secondaries (RU Peg, BV Cen and AE Aqr) it appeared that the dominant high latitude spot feature was located towards the trailing hemisphere --- raising the prospect that these large spots form at preferential longitudes. This does not appear to be the case for the high-latitude spot in either of the Roche tomograms of AE Aqr presented here.

There are two more prominent spot features in the Roche tomograms, particularly evident in figure~\ref{fig:block1tomogram}. The first is the large `appendage' extending from near the high-latitude spot at a latitude of $60^{\circ}$ down to $\sim10^{\circ}$ on the trailing hemisphere (labelled `A' in figure~\ref{fig:block1tomogram}). We note that this spot feature shows a distinct morphological change between the two observing runs. The second feature is the large spot near the rear of the star (labelled `B' in figure~\ref{fig:block1tomogram}). 

Finally, a `chain' of starspots is evident that lead down from $\sim 50^{\circ}$ latitude towards the L\tsb{1} point, and is offset towards the leading hemisphere by around $10\!-\!20^{\circ}$ in longitude. A similar chain of spots down to the L\tsb{1} point was seen on previous Roche tomograms of BV Cen and AE Aqr (\citealt{watson2007,watson2006}), and a higher spot coverage was seen on the hemisphere facing the white dwarf in the Doppler image of the pre-CV V471 Tau (\citealt{hussain2006}). This `chain' of spots suggests a mechanism which forces magnetic flux tubes to preferentially arise at these locations. This may be due to tidal forces, which are thought to be able to force spots to form at preferred longitudes (\citealt{holzwarth2003}), and may also be due to the tidal enhancement of the dynamo action itself (\citealt{moss2002}). In addition, surface flows may drag emergent magnetic flux tubes towards the L\tsb{1} point (discussed later in section~\ref{sec:surfaceflow}).

\begin{figure}
\centering
\includegraphics[width=0.5\textwidth]{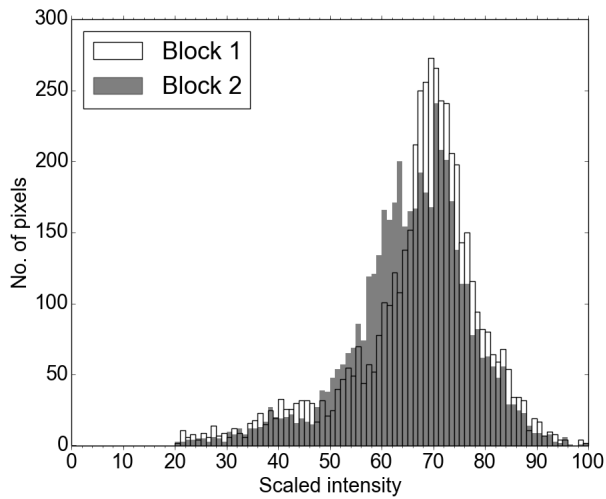}
\caption{Histograms of the pixel intensities from the Roche tomograms of block 1 and block 2 datasets of AE Aqr after pixels on the southern hemisphere (latitude $< 0^{\circ}$) were discarded. The brightest pixel in each map is assigned an intensity of 100 and other pixel intensities are scaled linearly against this. For the purposes of estimating the global spot properties of AE Aqr, the immaculate photosphere has been defined as pixels with intensities of 81 or greater, and the spotted photosphere those regions where pixel intensities are less than 81.}
\label{fig:histogram}
\end{figure}

In order to make a more quantitative estimate of the spot parameters on AE Aqr, we have examined the pixel intensity distribution in the Roche tomograms. We discarded all pixels in the southern hemisphere as, since this hemisphere is least visible, pixels are mainly assigned the default intensity, which in this case is the average intensity of the map. Histograms of the pixel values are shown in figure~\ref{fig:histogram} where we have assigned the brightest pixel in the map an intensity of 100 and linearly scaled the other pixel intensities relative to this. 

Unlike the previous Roche tomogram of AE Aqr by \cite{watson2006}, we do not see a clear bimodal distribution in pixel intensities from which we can confidently distinguish between immaculate photosphere and spotted photosphere. Instead, the histograms of pixel intensities show broad peaks, with long tails towards high and low pixel intensities. Pixels with a scaled intensity of around 50-64 may be explained by the smearing of spots across latitudes, increasing areal coverage and reducing contrast, and by unresolved spots which cause a decrease in pixel intensity without being clearly reconstructed in the tomograms. Above the peak in the histograms we see a tail of high intensity pixels. It is unlikely that these represent the immaculate photosphere, as the growth of bright pixels in maps which are not `threshholded' is a known artefact of Doppler imaging techniques (e.g. \citealt{hatzes1992} -- also see section~\ref{sec:rochetomography}). Instead, we have defined an intensity of 81 and above (the upper end of the peak of pixel intensities) to represent the immaculate photosphere.

We have defined a spot by taking the pixel intensity at the centre of the large, high-latitude spot to represent the intensity of a 100 per cent spotted region (in the same way as \citealt{watson2007}). Pixels with a lower intensity are present in the Roche tomogram, but these are confined to the irradiated region near the L\tsb{1} point. These are significantly lower in intensity, confirming our interpretation that this dark feature is not a spot.

If we assume a spot simply alters the continuum level and not line depths (a secondary influence on the LSD profiles), we can assume a blackbody scaling. This gives a temperature contrast between the photosphere and spot of $\Delta T = 570$ K for block 1 and $\Delta T = 640$ K for block 2. This is similar to $\Delta T = 780$ K found for BV Cen (see \citealt{watson2007}) and fits well with models by \cite{frasca2005} who found a temperature differences of $\Delta T = 450-850$ K, and \cite{biazzo2006} who found $\Delta T = 453-1012$ K for stars in various locations in the HR diagram. The small disagreement in the spot temperatures calculated for the two maps (and also the overall spot filling factors determined later) are most likely due to slight differences in the quality of the maps achieved. This is largely due to the technical fault which affected the block 1 data, which means that overall, small scale features are less readily reconstructed.


Each pixel in our Roche tomograms was given a spot-filling factor between 0 (immaculate) to 1 (totally spotted) depending on its intensity between our predefined immaculate and spotted photosphere intensities. After removal of the region near the L\tsb{1} point (which is irradiated and would cause us to overestimate the total spot coverage), we estimate that 15.4 per cent and 17 per cent of the northern hemisphere of AE Aqr is spotted, for block 1 and block 2 datasets, respectively. These values should be taken with caution since it depends on our classification of a spot, and is likely to be a lower limit due to the presence of unresolved spots, but they match a previous estimate of 18 per cent spot coverage by \cite{watson2006}, and is similar to the 22 per cent spot filling factor found by \cite{webb2002} in a TiO study of the CV SS Cyg.

\subsection{Spot coverage as a function of longitude and latitude}
In the analysis which follows, the map coordinates are defined such that $0^{\circ}$ longitude is the centre of the back of the star, with increasing longitude towards the leading hemisphere, and with the L\tsb{1} point at $180^{\circ}$. 

\begin{figure}
\centering
\includegraphics[width=0.5\textwidth]{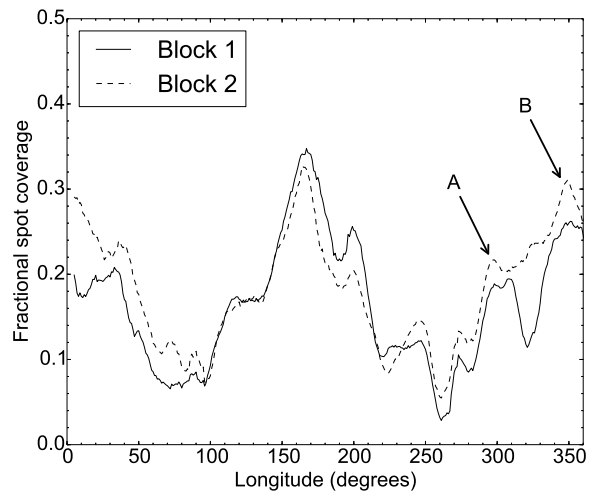}
\caption{Spot coverage on the northern hemisphere as a function of longitude for block 1 and block 2, normalised by the number of pixels within a 10 degree longitude bin. The region enclosed within $160-200^{\circ}$ longitude and $0-20^{\circ}$ latitude has been masked off for the L\tsb{1} point.}
\label{fig:spotcoveragelong}
\end{figure}

When plotted as a function of longitude (see figure~\ref{fig:spotcoveragelong}) we find a large fractional spot coverage around $40^{\circ},200^{\circ}, 310^{\circ}~\&~360^{\circ}$ longitude for both maps. The largest fractional coverage is between $150^{\circ}\!-\!210^{\circ}$ longitude. This longitude range contains the irradiated region around the L\tsb{1} point, but even without including this, the trail of spots down to the L\tsb{1} point have a large fractional spot coverage.

The two prominent features mentioned earlier, a large appendage labelled `A' and a prominent spot labelled `'B', are readily visible at longitudes of $300^{\circ}$ and $350^{\circ}$, respectively, in figure~\ref{fig:spotcoveragelong}. These become longitudinally blended in block 2 data due to differential rotation between maps.

\begin{figure}
\centering
\includegraphics[width=0.5\textwidth]{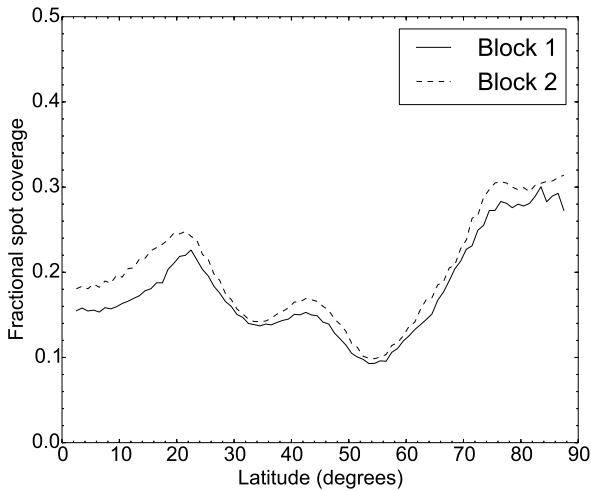}
\caption{Spot coverage on the northern hemisphere as a function of latitude for block 1 and block 2 data, normalised by the surface area at that latitude. Since the grid in our Roche tomograms is not aligned along strips of constant latitude, some interpolation between grid elements is required in order to produce these plots, resulting in their slightly noisy nature. The region around the L\tsb{1} point has been excluded.}
\label{fig:spotcoveragelat}
\end{figure}

As well as preferred longitudes, spots appear to form at preferred latitudes, as shown in figure~\ref{fig:spotcoveragelat}. There are two distinct bands around $22^{\circ}$ and $43^{\circ}$ latitude with a larger fractional spot coverage, in addition to the large spot coverage at latitudes above $70^{\circ}$. In the Roche tomogram of AE Aqr by \cite{watson2006} from observations in 2001, the authors reported evidence of increasing spot coverage towards lower latitudes, but their observations were not of sufficiently high signal-to-noise to determine whether these spots formed in a distinct latitude band. This work confirms that spots do indeed form at lower latitudes in distinct bands on AE Aqr. One key difference, however, between the 2001 map and the maps presented here is that we find a weak band of spots at a latitude centred around 43 degrees, whereas the 2001 observations show a distinct paucity of spots at the same latitude. This may be indicative of a solar-like magnetic activity cycle in operation in AE Aqr, where the latitude of spot formation may change in a manner that mimics the butterfly-diagram for the Sun. We tentatively speculate that the two latitude bands may be explained by a cross over between activity cycles, with the higher band just emerging and the lower band diminishing, as seen on the sun. The $\sim20^{\circ}$ band is similar to that found on BV Cen (\citealt{watson2007}).

The larger spots mentioned above are reconstructed independently on each tomogram, and are very similar in morphology (with some evolution). This similarity shows they are not artefacts due to noise or over fitting of the data.

\section{Differential rotation}
\label{sec:diffrot}
The Roche tomograms produced in section~\ref{sec:surfacemaps} were then used to measure the differential rotation rate on the secondary star. The maps are separated by 9 days, over which we expect minimal spot evolution, apart from changing morphology due to differential rotation. 

We assumed AE Aqr follows a solar-like differential rotation law of the form
\begin{equation}
\Omega(\theta) = \Omega_{eq} - d\Omega \sin^{2}\theta
\label{eqn:diffrotlaw}
\end{equation}
\noindent where $\Omega(\theta)$ is the rotation rate at latitude $\theta$, $\Omega_{eq}$ is the rotation rate at the equator, and $d\Omega$ is the difference between the rotation rates at the pole and the equator. 

Two different methods were used to measure the differential rotation rate, and for consistency both maps were reconstructed at an inclination of $51^{\circ}$, with $M_{1} = 1.17$ M$_{\odot}$ and $M_{2} = 0.78$ M$_{\odot}$. 

For the first method, we cross-correlated strips of constant latitude for each map from block 1 and block 2, taking the peak of the cross-correlation function (CCF) as the shift for that latitude. The tiling of the Roche lobe as implemented in Roche tomography, however, does not yield surface elements on strips of constant latitude, as this would lead to a discontinuity at the inner-Lagrangian point. For this reason we cannot incorporate a differential law into the reconstruction itself, as has been done by (for example) \cite{hussain2006} and \cite{petit2002}. Instead, we interpolated the maps using the `griddata' module from the SciPy library (\citealt{scipy}) to produce an equirectangular-projection with $1^{\circ} \times 1^{\circ}$ resolution, which was then used for the cross-correlations, as these interpolated maps now contain constant latitude strips.

In order to estimate the error on our derived differential rotation measurement, we adopted a Monte Carlo approach to assessing the impact of noise on the Roche tomography reconstructions. The usual method of `jiggling' the observed data in accordance with the error bars cannot be implemented in Roche tomography. This is because this process adds noise to the data, which means that the synthesized data sets are not being drawn from the same parent population as the observed data set and leads to maps dominated by noise. To surmount this issue, we took the reconstructed line profiles for each observation block and varied the flux of each data point about its value by the error bar in the observed data multiplied by a number output by a Gaussian random-number generator with zero mean and unit variance. In addition, to account for systematic effects in the observed data that were not reconstructed, we calculated a moving average for the residuals (thus removing unfitted noise from the observed data but keeping the large scale variation). This was added to the newly created line profiles, which were then visually checked to make sure the noise and shape of the line profiles were representative of the observed data. These new line profiles were then reconstructed to a similar map entropy as the observed data (the same reduced $\chi^2$ could not be used due to the variation introduced in the data). This whole process was repeated 100 times for each observation block and the resulting maps were cross-correlated as before, resulting in a distribution of shifts for each latitude. The uncertainty in the shifts found for the original maps was taken to be $3\sigma$ of this distribution.

\begin{figure}
\centering
\includegraphics[angle=0,width=0.5\textwidth]{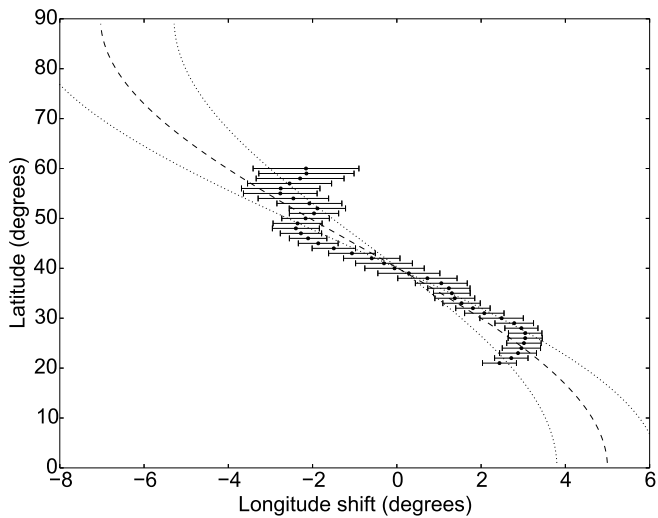}
\caption{The peaks of the cross-correlation of constant-latitude strips in the interpolated maps of block 1 and 2. The dashed line shows a minimised-$\chi^{2}$ best-fit of a solar-like differential rotation law, with $d\Omega = 0.00233$ rad d\tsu{-1} and a co-rotation latitude of $40.1^{\circ}$. The dotted lines show $chi^{2} + 1$ fits, with a minimum and maximum $d\Omega$ of 0.0176 rad d\tsu{-1} and 0.0291 rad d\tsu{-1}, respectively.}
\label{fig:ccfit}
\end{figure}

The CCFs for each latitude may be found in an additional figure online. It was found that below $20^{\circ}$ latitude no significant shift was found, most likely due to the broad CCF peaks stemming from a lack of well defined features on the map. Note, the region $\pm20^\circ$ around the L\tsb{1} point up to $20^\circ$ latitude was ignored in the CCF due to the large, unchanging irradiation feature. Additionally, no shift was found above $60^{\circ}$ latitude, where the broad CCF peaks stem from the large polar-region spot appearing unchanged between block 1 and 2. For these reasons, we only included latitudes between $20\!-\!60^{\circ}$ in our analysis. A minimised-$\chi^{2}$ fit of equation~\ref{eqn:diffrotlaw} was made to the CCF peaks, iteratively adding an offset to account for the co-rotation latitude. This gave $d\Omega = 0.0233$ rad d\tsu{-1} (corresponding to a equator-pole lap time of 269 days) and a co-rotation latitude of $40.1^{\circ}$. The $\chi^{2} + 1$ limits of this fit gave $d\Omega = 0.0176$ rad d\tsu{-1} and $d\Omega = 0.0291$ rad d\tsu{-1}, with co-rotation latitudes of $40.3^{\circ}$ and $40.1^{\circ}$, respectively (see figure~\ref{fig:ccfit}).

\begin{figure}
\centering
\includegraphics[width=0.5\textwidth]{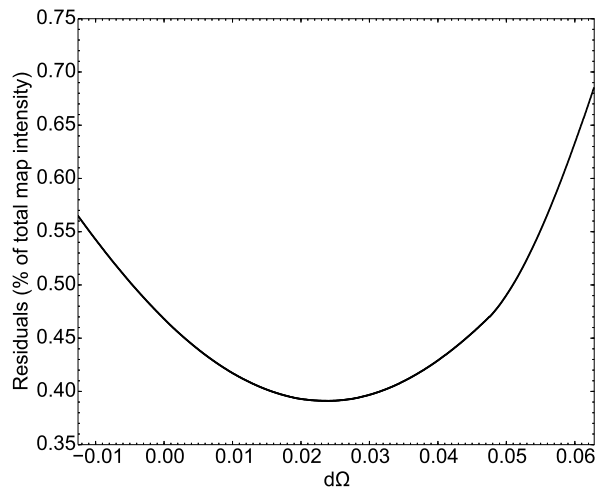}
\caption{The residuals after applying a solar-like DR law to the interpolated block 1 data and subtracting the interpolated block 2 data (see text for full description). The minimum gives $d\Omega = 0.02396$ rad d\tsu{-1} and a co-rotation latitude of $41.8^{\circ}$.}
\label{fig:interpolatedresiduals}
\end{figure}

In addition, to check the results from the CCF analysis, we implemented a second approach to measure the shear rate in the tomograms. This consisted of applying different levels of differential rotation to the 1\tsu{st} map (reconstructed from block 1 data and assuming a solar-like differential law shown in equation~\ref{eqn:diffrotlaw}), and subtracting this `sheared map' from the second map and analysing the residuals. In this process, the level of applied shear that results in the least residuals is deemed to best represent the true differential rotation rate. As before, since this requires strips of constant latitude, we used the map interpolated using the `griddata' module as described earlier. The residuals found after map subtraction were weighted by $\cos{\theta}$ (where $\theta$ is the latitude) since the map projection increases the weighting of higher latitudes if not corrected for. The residuals were then squared and summed and are shown in figure~\ref{fig:interpolatedresiduals} as a percentage of the summed pixel intensities in the interpolated map of block 1 data. The least-residuals fit yielded $d\Omega = 0.02396$ rad d\tsu{-1} (corresponding to a equator-pole lap time of 262 days) and a co-rotation latitude of $41.8^{\circ}$, in good agreement with the CCF analysis.

\begin{figure}
\centering
\includegraphics[width=0.5\textwidth]{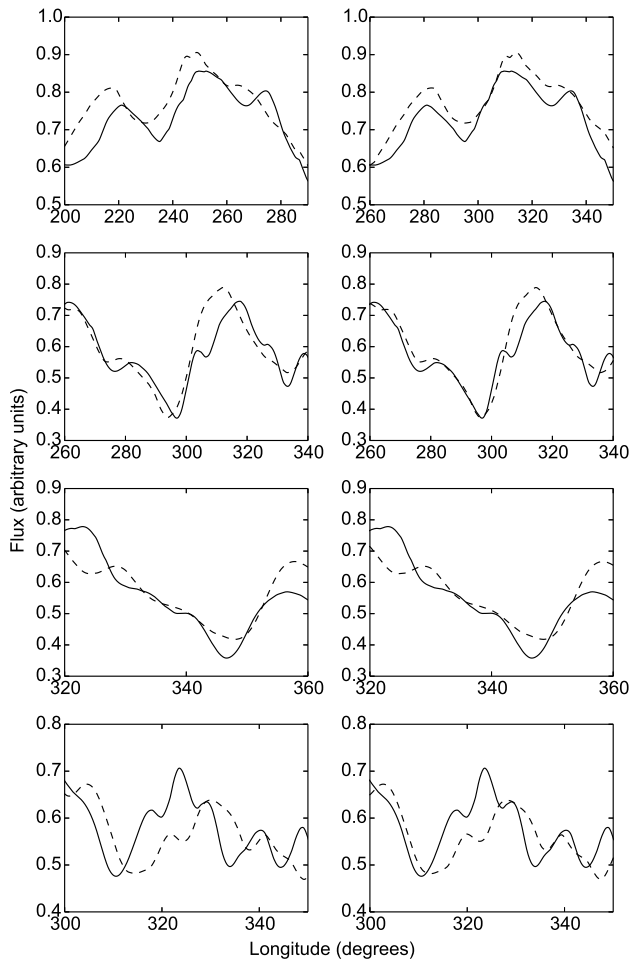}
\caption{The map intensity as a function of longitude. The solid line is block 1 data and the dashed line is block 2 data. The left column shows the unshifted data and the right column shows the same data after applying the shift found from the fit to the CCF peaks (see figure~\ref{fig:ccfit}). The panels from top to bottom are: a bright region at $68^{\circ}$ latitude; the top of appendage 'A' at $52^{\circ}$ latitude; a prominent spot 'B' at $42^{\circ}$ latitude; the bottom of appendage 'A' at $30^{\circ}$. Note, the second and fourth panels down on the right are visually not shifted enough, reflecting the fit shown in figure~\ref{fig:ccfit}.}
\label{fig:snippets}
\end{figure}

To more clearly show the movement of surface features, segments of the constant latitude strips are shown in figure~\ref{fig:snippets}, showing the surface intensity across the longitude range for various regions of the maps.

This analysis was also carried out after varying the masses by $\pm10$ per cent, yielding equator-pole lap times of 278-354 d by changing $M_{2}$, and 336-345 d by changing $M_{1}$. However, these are not representative of the uncertainty on the shear rate since incorrect masses cause large artefacts to dominate the maps, thus diluting the contribution made by real features. To provide a more realistic error estimate, the systemic velocity was varied by $\pm1$ kms\tsu{-1}, yielding a lap time of 267-296 d. The analysis was also carried out at an inclination of $66^{\circ}$ using $M_{1} = 0.74$ M$_{\odot}$ and $M_{2} = 0.5$ M$_{\odot}$, as found by Watson et al. \citeyearpar{watson2006}. This yielded equator-pole lap times of 322 d and 316 d and co-rotation latitudes of $39.5^{\circ}$ and $40.7^{\circ}$ for the first and second methods, respectively. These are similar to the results found above, and show the robustness of the shear measurement against an incorrect inclination and systemic velocity. We therefore estimate the shear rate uncertainty to be $d\Omega = 0.0233 \pm0.0057$ rad d\tsu{-1} and the corresponding lap time to be $269^{+88}_{-53}$ d.

\section{Discussion}
\label{sec:discussion}
The measurement of differential rotation (DR) in a CV donor is the first of its kind and shows that the secondary star is (technically) not tidally locked. Theoretically, DR is predicted to be weak in tidally distorted systems (\citealt{scharlemann1982}) and the result found in this work confirms this observationally.

When compared to other systems we find the DR rate determined in this work is much lower than that for single stars, although LQ Hya is very similar, despite its rotation being four times slower. The pre-CV V471 Tau was found by \cite{hussain2006} to have a drastically different DR rate (essentially that of a solid body), despite having a similar rotation period and mass, suggesting that the filling of the star's Roche lobe may impact the DR rate. Finally, HR 1099 is significantly different in every parameter. The lack of systems with similar parameters with which to compare means no significant conclusions can be drawn, although we \emph{can} say DR rates vary wildly in the systems observed so far.

\begin{table*}
\caption{Comparison of system types and differential rotation rates, as found by the respective authors.}
\label{tab:diffrotcompare}
\begin{tabular}{ccccccc}
\toprule
Author & Object & System type & Stellar type & $P_{rot}$ (hrs) & $M_{2}$ (M$_{\odot}$) & Lap time (days)\\
\midrule
This work&AE Aqr	&CV secondary	&K4V	&9.88	&0.78-0.8	&269 \\
(cross-correlation method)&&&&&&\\
This work&AE Aqr	&CV secondary	&K4V	&9.88	&0.78-0.8	&262 \\
(map subtraction method)&&&&&&\\
\cite{hussain2006}	&V471 Tau	&Tidally-locked pre-CV	&K2V	&12.5	&$0.79\pm0.04$	&3900 \\
\cite{petit2004}	&HR 1099	&Tidally locked RS CVn	&K1	&68.1	&$1.1\pm0.2$	&480 \\
\cite{cameron2001}	&AB Dor	&Single star&K0V	&12.36	&0.76	&70-140 \\
&&(in a distant trinary system)&&&&\\
\cite{donati2000}	&RX J$1508.6\!\pm\!4423$	&Single star	&G2	&7.4	&$1.16\pm0.04$	&$50\pm10$ \\
\cite{barnes2000}	&PZ Tel	&Single star	&G9IV/V	&22.68	&1.1	&$72\!-\!100$ \\
\cite{kovari2005}	&LQ Hya	&Single star	&K2	&38.4	&0.8	&280 \\
\bottomrule
\end{tabular}
(3)~$M_{2}$ -- \cite{guinan2001}; (4)~$M_{2}$ -- \cite{fekel1983}
\end{table*}

\subsection{Mass transfer variations}
\label{sec:masstranvar}
The presence of differential rotation on AE Aqr has significant implications for theories invoking starspots to explain rapid variations in accretion rates. The passage of spots across the mass-transfer nozzle may cease mass-transfer due to the depression of the photosphere by strong magnetic fields in a starspot, decreasing thermal pressure and density (\citealt{livio1994}). \cite{hessman2000}, in their study of mass-transfer variations in AM Her, concluded that about half the surface of the star near the L\tsb{1} point needs to be covered with spots to produce the observed variations. They note this scenario requires a mechanism which preferentially produces magnetic flux around the L\tsb{1} point and/or one which forces spot groups, which appear at higher latitudes, to wander down towards the L\tsb{1} point. The trail of spots leading down to the L\tsb{1} point, which is evident in several previous Roche tomograms as well as this work, lends observational support for this scenario (for further discussion see section~\ref{sec:surfaceflow}).

\cite{king1998} note that the most rapid transitions to low states occur over the course of a day. Using our measurement of differential rotation, we can estimate how long a spot would take to traverse the mass-transfer nozzle and cut off mass-transfer. We estimate the mass-transfer nozzle radius of AE Aqr to be $\sim16,000$ kms\tsu{-1} (see section 4 of \citealt{hessman2000}). Taking the average radius of the star as $8.4\times10^{5}$ km, combined with the $\sim5^{\circ}$ equatorial longitude-shift over the 9 day observation gap, we estimate the equator rotates at a speed of $\sim90$ ms\tsu{-1}. This means it takes $\sim4$ days for a spot to fully cover the mass-transfer nozzle, assuming the spotted region is large enough, and that the surface flow is constant across the L\tsb{1} region. This is similar to the length of time taken for the large dips in brightness observed in AM Her, which has a much shorter orbital period and is of a different spectral type, suggesting that this is a viable mechanism.

In addition, the highly magnetised region around the mass-transfer nozzle may cause the formation of magnetised blobs of material in the accretion stream, causing the flicker seen in CV lightcurves (\citealt{bruch1992}).

\subsection{Surface flows: Impact on starspot movement}
\label{sec:surfaceflow}
To understand why surface features are located where they are, and to determine a mechanism which preferentially produces magnetic flux around the L\tsb{1} point, we can compare the Roche tomograms to theoretical predictions and numerical simulations of surface mass flow on Roche lobe filling secondaries. 

\cite{oka2002} define the `surface' as the region where appreciable mass flow occurs, and is not necessarily the photosphere. \cite{lubow1975} state that mass flow is directed around contours of equal pressure, proceeding from high to low pressure, with the highest pressure at the poles. They predict that surface flow may be astrostrophic (i.e. parallel to isobars) on each equipotential surface, but that matter may be transferred from the entire surface layer to the equator if there is an outward flux at the bottom of the surface layer. 

In their numerical simulations, \cite{oka2002} find gas ascends in higher density regions, and has a circulating flow in lower density regions. They find a high pressure circulatory flow around the north pole (denoted the H-eddy), where gas seems to drift gradually downward, increasing its velocity with decreasing latitude. Near the L\tsb{1} region they find a strong low pressure, due to outflow through the L\tsb{1} point, and a circulatory flow around the L\tsb{1} point. In a counter-cockwise rotating secondary (as viewed from the north pole), this L\tsb{1} eddy rotates in a coulter-clockwise direction. The L\tsb{1} eddy is fixed to the L\tsb{1} region, but as this forces the H-eddy towards leading hemisphere, the system is not steady. To stabilise the eddy system, a third circulation is needed in the form of a L2 eddy on the back of the star, which circulates counter-clockwise and acts to convect the H-eddy towards the trailing hemisphere direction. These three eddy features are present in simulations covering a mass-ratio range $q = 0.05 - 3$ (\citealt{oka2004}).


These simulations may provide an explanation for the trail of spots leading from the pole to the L\tsb{1} point, as seen in Roche tomograms of AE Aqr and BV Cen (\citealt{watson2006,watson2007}; this work). These may be caused by the combined effect of the H-eddy and the L\tsb{1} eddy, bringing magnetic flux tubes which emerge near polar regions down the leading hemisphere side of the L\tsb{1} region. In addition, a combination of the H-eddy and L2 eddy may cause magnetic flux tubes to be dragged down the trailing hemisphere, creating the large appendage we see in the AE Aqr tomograms in this work, and additionally causing the morphological change we see between the two blocks. This would lead to a skewed measure of differential rotation, since the surface would be circulating rather than truly rotating longitudinally. However, this explanation is lacking as the simulations by Oka et al. do not account for magnetic effects, so we do not know how magnetic flux tubes are influenced by this ``surface" flow.



\subsection{DR rate variations and future opportunities}
Substantial changes in differential rotation on magnetically active binaries may cause quasi-cyclic orbital period changes (\citealt{applegate1992}). A stellar activity cycle may alter the viscous transport of angular momentum sufficiently to produce substantial changes in DR, altering the gravitational quadropole moment of the star sufficiently to produce orbital period changes (e.g. \citealt{donati1999}). \cite{cameron2002abdor} suggest that the large modulation of surface DR in AB Dor would alter the stars oblatness such that if it were a close binary system it would be expected to produce observable long-term orbital period changes. This effect could be tested in AE Aqr with future DR measurements, testing if the Applegate mechanism exists in a Roche lobe filling star. In addition, variation in DR could alter the mass transfer rate (see section~\ref{sec:masstranvar}) such that it would affect the evolution of the binary system as a whole.
This is the first measure of DR for a CV and it would be interesting to compare this to another with different system parameters. A good candidate is the secondary in BV Cen, as it is Roche lobe filling, has a longer orbital period ($P_{orb} = 14.67$ hrs) and is more massive ($M_2 = 1.05 \mathrm{M}_{\odot}$). We intend to carry out this study in the near future.


\section{Conclusions}
We have unambiguously imaged starspots on two Roche tomograms of AE Aqr, separated by 9 days. We find a spot coverage of $15.4-17$ per cent. A comparison with a previous Roche tomogram of AE Aqr by \cite{watson2006} shows many similarities, confirming the highly spotted nature of the secondary. The spot distributions indicate preferred latitudes for spot formation, with a `chain' of spots from the pole to the L\tsb{1} point and the two distinct latitude bands of spots around $22^{\circ}$ and $43^{\circ}$ latitude. 

Using these two Roche tomograms, we have measured the surface differential rotation using two different techniques --- cross-correlation of constant-latitude strips, and subtraction of maps after injecting a solar-like differential law into the first tomogram. The first yields an equator-pole lap time of 269 days and the second yields a lap time of 262 days, with a co-rotation latitude of $\sim40^{\circ}$. This shows that the star is not tidally locked (as was previously assumed for CVs), and is in stark contrast with the near solid body rotation of the tidally locked pre-CV V471 Tau, found by \cite{hussain2006}. The discovery of differential rotation has large implications for stellar dynamo theory, and the cessation of mass transfer due to spot traversal of the L\tsb{1} point. The highly magnetised region around the L\tsb{1} point may create magnetised blobs of material, explaining the `blobby' accretion observed in CVs.

\section*{Acknowledgments}
We thank Tom Marsh for the use of his \textsc{molly} software package in this work, and VALD for the stellar line-lists used. C.A.H. acknowledges the Queen's University Belfast Department of Education and Learning PhD scholarship, C.A.W. acknowledges support by STFC grant ST/L000709/1. T.S. acknowledges support by the Spanish Ministry of Economy and Competitiveness (MINECO) under the grant (project reference AYA2010-18080). This research has made use of NASA's Astrophysics Data System and the Ureka software package provided by Space Telescope Science Institute and Gemini Observatory. Finally, we would like to thank the referee, Gaitee Hussain, for their comments that helped to substantially improve the paper.

\bibliographystyle{mn_new}
\bibliography{references}

\begin{thebibliography}{71}
\expandafter\ifx\csname natexlab\endcsname\relax\def\natexlab#1{#1}\fi

\bibitem[{{Ak} et~al.(2001){Ak}, {Ozkan}, \& {Mattei}}]{ak2001}
{Ak}, T., {Ozkan}, M.~T., {Mattei}, J.~A., 2001, \aap, 369, 882

\bibitem[{{Applegate}(1992)}]{applegate1992}
{Applegate}, J.~H., 1992, \apj, 385, 621

\bibitem[{{Barnes}(1999)}]{barnes1999}
{Barnes}, J., 1999, PhD thesis, University of St. Andrews

\bibitem[{{Barnes} et~al.(2000){Barnes}, {Collier Cameron}, {James}, \&
  {Donati}}]{barnes2000}
{Barnes}, J.~R., {Collier Cameron}, A., {James}, D.~J., {Donati}, J.-F., 2000,
  \mnras, 314, 162

\bibitem[{{Barnes} et~al.(2004){Barnes}, {Lister}, {Hilditch}, \& {Collier
  Cameron}}]{barnes2004}
{Barnes}, J.~R., {Lister}, T.~A., {Hilditch}, R.~W., {Collier Cameron}, A.,
  2004, \mnras, 348, 1321

\bibitem[{{Bianchini}(1990)}]{bianchini1990}
{Bianchini}, A., 1990, \aj, 99, 1941

\bibitem[{{Biazzo} et~al.(2006){Biazzo}, {Frasca}, {Catalano}, {Marilli},
  {Henry}, \& {T{\v a}s}}]{biazzo2006}
{Biazzo}, K., {Frasca}, A., {Catalano}, S., {Marilli}, E., {Henry}, G.~W.,
  {T{\v a}s}, G., 2006, Memorie della Societa Astronomica Italiana Supplementi,
  9, 220

\bibitem[{{Bruch}(1991)}]{bruch1991}
{Bruch}, A., 1991, \aap, 251, 59

\bibitem[{{Bruch}(1992)}]{bruch1992}
{Bruch}, A., 1992, \aap, 266, 237

\bibitem[{{Cameron}(2001)}]{cameron2001}
{Cameron}, A.~C., 2001, in {H.~M.~J.~Boffin, D.~Steeghs, \& J.~Cuypers}, ed.,
  Astrotomography, Indirect Imaging Methods in Observational Astronomy, vol.
  573 of \emph{Lecture Notes in Physics, Berlin Springer Verlag}, p. 183

\bibitem[{{Casares} et~al.(1996){Casares}, {Mouchet}, {Martinez-Pais}, \&
  {Harlaftis}}]{casares1996}
{Casares}, J., {Mouchet}, M., {Martinez-Pais}, I.~G., {Harlaftis}, E.~T., 1996,
  \mnras, 282, 182

\bibitem[{{Chanan} et~al.(1976){Chanan}, {Middleditch}, \&
  {Nelson}}]{chanan1976}
{Chanan}, G.~A., {Middleditch}, J., {Nelson}, J.~E., 1976, \apj, 208, 512

\bibitem[{{Chincarini} \& {Walker}(1981)}]{chincarini1981}
{Chincarini}, G., {Walker}, M.~F., 1981, \aap, 104, 24

\bibitem[{{Claret}(2000)}]{claret2000}
{Claret}, A., 2000, VizieR Online Data Catalog, 336, 31081

\bibitem[{{Collier Cameron} \& {Donati}(2002)}]{cameron2002abdor}
{Collier Cameron}, A., {Donati}, J.-F., 2002, \mnras, 329, L23

\bibitem[{{Collier-Cameron} \& {Unruh}(1994)}]{cameron1994}
{Collier-Cameron}, A., {Unruh}, Y.~C., 1994, \mnras, 269, 814

\bibitem[{{Crawford} \& {Kraft}(1956)}]{crawford1956}
{Crawford}, J.~A., {Kraft}, R.~P., 1956, \apj, 123, 44

\bibitem[{{Davey} \& {Smith}(1992)}]{davey1992}
{Davey}, S., {Smith}, R.~C., 1992, \mnras, 257, 476

\bibitem[{{Dhillon} \& {Watson}(2001)}]{dhillon2001}
{Dhillon}, V.~S., {Watson}, C.~A., 2001, in {H.~M.~J.~Boffin, D.~Steeghs, \&
  J.~Cuypers}, ed., Astrotomography, Indirect Imaging Methods in Observational
  Astronomy, vol. 573 of \emph{Lecture Notes in Physics, Berlin Springer
  Verlag}, p.~94

\bibitem[{{Donati}(1999)}]{donati1999}
{Donati}, J.-F., 1999, \mnras, 302, 457

\bibitem[{{Donati} \& {Collier Cameron}(1997)}]{donati1997abdor}
{Donati}, J.-F., {Collier Cameron}, A., 1997, \mnras, 291, 1

\bibitem[{{Donati} et~al.(1997){Donati}, {Semel}, {Carter}, {Rees}, \& {Collier
  Cameron}}]{donati1997}
{Donati}, J.-F., {Semel}, M., {Carter}, B.~D., {Rees}, D.~E., {Collier
  Cameron}, A., 1997, \mnras, 291, 658

\bibitem[{{Donati} et~al.(2000){Donati}, {Mengel}, {Carter}, {Marsden},
  {Collier Cameron}, \& {Wichmann}}]{donati2000}
{Donati}, J.-F., {Mengel}, M., {Carter}, B.~D., {Marsden}, S., {Collier
  Cameron}, A., {Wichmann}, R., 2000, \mnras, 316, 699

\bibitem[{{Dunford} et~al.(2012){Dunford}, {Watson}, \& {Smith}}]{dunford2012}
{Dunford}, A., {Watson}, C.~A., {Smith}, R.~C., 2012, \mnras, 422, 3444

\bibitem[{{Echevarr{\'{\i}}a} et~al.(2008){Echevarr{\'{\i}}a}, {Smith},
  {Costero}, {Zharikov}, \& {Michel}}]{echevarria2008}
{Echevarr{\'{\i}}a}, J., {Smith}, R.~C., {Costero}, R., {Zharikov}, S.,
  {Michel}, R., 2008, \mnras, 387, 1563

\bibitem[{Efron(1979)}]{efron1979}
Efron, B., 1979, The Annals of Statistics, 7, 1

\bibitem[{Efron \& Tibshirani(1993)}]{efron1993}
Efron, B., Tibshirani, R., 1993, An introduction to the bootstrap, Monographs
  on statistics and applied probability, Chapman \& Hall

\bibitem[{{Fekel}(1983)}]{fekel1983}
{Fekel}, Jr., F.~C., 1983, \apj, 268, 274

\bibitem[{{Frasca} et~al.(2005){Frasca}, {Biazzo}, {Catalano}, {Marilli},
  {Messina}, \& {Rodon{\`o}}}]{frasca2005}
{Frasca}, A., {Biazzo}, K., {Catalano}, S., {Marilli}, E., {Messina}, S.,
  {Rodon{\`o}}, M., 2005, \aap, 432, 647

\bibitem[{{Guinan} \& {Ribas}(2001)}]{guinan2001}
{Guinan}, E.~F., {Ribas}, I., 2001, \apjl, 546, L43

\bibitem[{{Hatzes} \& {Vogt}(1992)}]{hatzes1992}
{Hatzes}, A.~P., {Vogt}, S.~S., 1992, \mnras, 258, 387

\bibitem[{{Hessman} et~al.(2000){Hessman}, {G{\"a}nsicke}, \&
  {Mattei}}]{hessman2000}
{Hessman}, F.~V., {G{\"a}nsicke}, B.~T., {Mattei}, J.~A., 2000, \aap, 361, 952

\bibitem[{{Holzwarth} \& {Sch{\"u}ssler}(2003)}]{holzwarth2003}
{Holzwarth}, V., {Sch{\"u}ssler}, M., 2003, \aap, 405, 303

\bibitem[{{Hussain} et~al.(2006){Hussain}, {Allende Prieto}, {Saar}, \&
  {Still}}]{hussain2006}
{Hussain}, G.~A.~J., {Allende Prieto}, C., {Saar}, S.~H., {Still}, M., 2006,
  \mnras, 367, 1699

\bibitem[{{Hussain} et~al.(2007)}]{hussain2007}
{Hussain}, G.~A.~J., et~al., 2007, \mnras, 377, 1488

\bibitem[{{Jeffers} et~al.(2002){Jeffers}, {Barnes}, \& {Collier
  Cameron}}]{jeffers2002}
{Jeffers}, S.~V., {Barnes}, J.~R., {Collier Cameron}, A., 2002, \mnras, 331,
  666

\bibitem[{{K{\H o}v{\'a}ri} et~al.(2005){K{\H o}v{\'a}ri}, {Weber}, \&
  {Strassmeier}}]{kovari2005}
{K{\H o}v{\'a}ri}, Z., {Weber}, M., {Strassmeier}, K.~G., 2005, in {Favata},
  F., {Hussain}, G.~A.~J., {Battrick}, B., eds., 13th Cambridge Workshop on
  Cool Stars, Stellar Systems and the Sun, vol. 560 of \emph{ESA Special
  Publication}, p. 731

\bibitem[{{King} \& {Cannizzo}(1998)}]{king1998}
{King}, A.~R., {Cannizzo}, J.~K., 1998, \apj, 499, 348

\bibitem[{{Kraft}(1967)}]{kraft1967}
{Kraft}, R.~P., 1967, \apj, 150, 551

\bibitem[{{Kupka} \& {Ryabchikova}(1999)}]{kupka1999}
{Kupka}, F., {Ryabchikova}, T.~A., 1999, Publications de l'Observatoire
  Astronomique de Beograd, 65, 223

\bibitem[{{Kupka} et~al.(2000){Kupka}, {Ryabchikova}, {Piskunov}, {Stempels},
  \& {Weiss}}]{kupka2000}
{Kupka}, F.~G., {Ryabchikova}, T.~A., {Piskunov}, N.~E., {Stempels}, H.~C.,
  {Weiss}, W.~W., 2000, Baltic Astronomy, 9, 590

\bibitem[{{Livio} \& {Pringle}(1994)}]{livio1994}
{Livio}, M., {Pringle}, J.~E., 1994, \apj, 427, 956

\bibitem[{{Lubow} \& {Shu}(1975)}]{lubow1975}
{Lubow}, S.~H., {Shu}, F.~H., 1975, \apj, 198, 383

\bibitem[{{Marsden} et~al.(2005){Marsden}, {Waite}, {Carter}, \&
  {Donati}}]{marsden2005}
{Marsden}, S.~C., {Waite}, I.~A., {Carter}, B.~D., {Donati}, J.-F., 2005,
  \mnras, 359, 711

\bibitem[{{Mestel}(1968)}]{mestel1968}
{Mestel}, L., 1968, \mnras, 138, 359

\bibitem[{{Moss} et~al.(2002){Moss}, {Piskunov}, \& {Sokoloff}}]{moss2002}
{Moss}, D., {Piskunov}, N., {Sokoloff}, D., 2002, \aap, 396, 885

\bibitem[{{Oka} et~al.(2002){Oka}, {Nagae}, {Matsuda}, {Fujiwara}, \&
  {Boffin}}]{oka2002}
{Oka}, K., {Nagae}, T., {Matsuda}, T., {Fujiwara}, H., {Boffin}, H.~M.~J.,
  2002, \aap, 394, 115

\bibitem[{{Oka} et~al.(2004){Oka}, {Matsuda}, {Hachisu}, \& {Boffin}}]{oka2004}
{Oka}, K., {Matsuda}, T., {Hachisu}, I., {Boffin}, H.~M.~J., 2004, \aap, 419,
  277

\bibitem[{Oliphant(2007)}]{scipy}
Oliphant, T.~E., 2007, Computing in Science \& Engineering, 9, 10

\bibitem[{{Petit} et~al.(2002){Petit}, {Donati}, \& {Collier
  Cameron}}]{petit2002}
{Petit}, P., {Donati}, J.-F., {Collier Cameron}, A., 2002, \mnras, 334, 374

\bibitem[{{Petit} et~al.(2004)}]{petit2004}
{Petit}, P., et~al., 2004, in {A.~Maeder \& P.~Eenens}, ed., Stellar Rotation,
  vol. 215 of \emph{IAU Symposium}, p. 294

\bibitem[{{Rappaport} et~al.(1983){Rappaport}, {Verbunt}, \&
  {Joss}}]{rappaport1983}
{Rappaport}, S., {Verbunt}, F., {Joss}, P.~C., 1983, \apj, 275, 713

\bibitem[{{Richman} et~al.(1994){Richman}, {Applegate}, \&
  {Patterson}}]{richman1994}
{Richman}, H.~R., {Applegate}, J.~H., {Patterson}, J., 1994, Publications of
  the Astronomical Society of the Pacific, 106, 1075

\bibitem[{{Rutten} \& {Dhillon}(1994)}]{rutten1994}
{Rutten}, R.~G.~M., {Dhillon}, V.~S., 1994, \aap, 288, 773

\bibitem[{{Rutten} \& {Dhillon}(1996)}]{rutten1996}
{Rutten}, R.~G.~M., {Dhillon}, V.~S., 1996, in {A.~Evans \& J.~H.~Wood}, ed.,
  IAU Colloq. 158: Cataclysmic Variables and Related Objects, vol. 208 of
  \emph{Astrophysics and Space Science Library}, p.~21

\bibitem[{{Scharlemann}(1982)}]{scharlemann1982}
{Scharlemann}, E.~T., 1982, \apj, 253, 298

\bibitem[{{Schuessler} \& {Solanki}(1992)}]{schuessler1992}
{Schuessler}, M., {Solanki}, S.~K., 1992, \aap, 264, L13

\bibitem[{{Schwope} et~al.(2004){Schwope}, {Staude}, {Vogel}, \&
  {Schwarz}}]{schwope2004}
{Schwope}, A.~D., {Staude}, A., {Vogel}, J., {Schwarz}, R., 2004, Astronomische
  Nachrichten, 325, 197

\bibitem[{{Shahbaz} et~al.(2014){Shahbaz}, {Watson}, \&
  {Dhillon}}]{shahbaz2014}
{Shahbaz}, T., {Watson}, C.~A., {Dhillon}, V.~S., 2014, \mnras

\bibitem[{{Simpson} et~al.(2011)}]{simpson2011}
{Simpson}, E.~K., et~al., 2011, \mnras, 414, 3023

\bibitem[{{Skilling} \& {Bryan}(1984)}]{skilling1984}
{Skilling}, J., {Bryan}, R.~K., 1984, \mnras, 211, 111

\bibitem[{{Spruit} \& {Ritter}(1983)}]{spruit1983}
{Spruit}, H.~C., {Ritter}, H., 1983, \aap, 124, 267

\bibitem[{{Tanzi} et~al.(1981){Tanzi}, {Chincarini}, \& {Tarenghi}}]{tanzi1981}
{Tanzi}, E.~G., {Chincarini}, G., {Tarenghi}, M., 1981, \pasp, 93, 68

\bibitem[{{Vogt} \& {Penrod}(1983)}]{vogt1983}
{Vogt}, S.~S., {Penrod}, G.~D., 1983, Publications of the Astronomical Society
  of the Pacific, 95, 565

\bibitem[{{Vogt} et~al.(1999){Vogt}, {Hatzes}, {Misch}, \&
  {K{\"u}rster}}]{vogt1999}
{Vogt}, S.~S., {Hatzes}, A.~P., {Misch}, A.~A., {K{\"u}rster}, M., 1999, The
  \apj Supplement Series, 121, 547

\bibitem[{{Watson} \& {Dhillon}(2001)}]{watson2001}
{Watson}, C.~A., {Dhillon}, V.~S., 2001, \mnras, 326, 67

\bibitem[{{Watson} et~al.(2003){Watson}, {Dhillon}, {Rutten}, \&
  {Schwope}}]{watson2003}
{Watson}, C.~A., {Dhillon}, V.~S., {Rutten}, R.~G.~M., {Schwope}, A.~D., 2003,
  \mnras, 341, 129

\bibitem[{{Watson} et~al.(2006){Watson}, {Dhillon}, \& {Shahbaz}}]{watson2006}
{Watson}, C.~A., {Dhillon}, V.~S., {Shahbaz}, T., 2006, \mnras, 368, 637

\bibitem[{{Watson} et~al.(2007){Watson}, {Steeghs}, {Shahbaz}, \&
  {Dhillon}}]{watson2007}
{Watson}, C.~A., {Steeghs}, D., {Shahbaz}, T., {Dhillon}, V.~S., 2007, \mnras,
  382, 1105

\bibitem[{{Webb} et~al.(2002){Webb}, {Naylor}, \& {Jeffries}}]{webb2002}
{Webb}, N.~A., {Naylor}, T., {Jeffries}, R.~D., 2002, The \apj, 568, L45

\bibitem[{{Welsh} et~al.(1995){Welsh}, {Horne}, \& {Gomer}}]{welsh1995}
{Welsh}, W.~F., {Horne}, K., {Gomer}, R., 1995, \mnras, 275, 649

\end{thebibliography}
\label{lastpage}

\end{document}